\documentclass[twoside]{article}

\usepackage[numbers, compress]{natbib}
\usepackage{aistats2020}
\usepackage{hyperref}       % hyperlinks
\usepackage{url}            % simple URL typesetting
\usepackage{booktabs}       % professional-quality tables
\usepackage{amsfonts}       % blackboard math symbols
\usepackage{nicefrac}       % compact symbols for 1/2, etc.
\usepackage{microtype}      % microtypography
\usepackage{graphicx}
\usepackage{algorithm, algorithmic}
\usepackage{amsmath, layout, verbatim,rotating,enumerate,amssymb,amsthm,subfig, url}

\renewcommand{\algorithmicrequire}{\textbf{\small Input:}}
\renewcommand{\algorithmicensure}{\textbf{\small Output:}}
\usepackage{color}
\def\X{***ATT***}
\renewcommand{\comment}[1]{}

\definecolor{lightgray}{gray}{0.6}

\usepackage{xcolor}
\definecolor{darkred}{HTML}{660022}

\newcommand{\add}[1]{{\color{black} #1}} %ND: changed to black

\newcommand{\edit}[1]{{\color{black} #1}} %ND: changed to black

\renewcommand{\vec}[1]{\mathbf{#1}}
\renewcommand{\hat}[1]{\widehat{#1}}

\def\L{{\mathcal L}}

\def\P{{\mathbb P}}
\def\E{{\mathbb E}}

\def\X{{\vec{X}}}
\def\x{{\vec{x}}}

\newcommand{\mP}{\mathbb{P}}
\newcommand{\mE}{\mathbb{E}}

\newtheorem{thm}{Theorem}
\newtheorem{Def}{Definition}
\newtheorem{Cor}{Corollary}

\newtheorem{Lemma}{Lemma}
\newtheorem{remark}{Remark}
\newtheorem{Assumption}{Assumption}

\newcommand{\bx}{\mathbf{x}}
\newcommand{\bX}{\textbf{X}}

\begin{document}



\twocolumn[

\aistatstitle{Validation of Approximate Likelihood and Emulator Models for Computationally Intensive Simulations}

%\aistatsauthor{Anonymous Authors}
%\aistatsaddress{}
%]
\vspace{-0.25cm}
\aistatsauthor{Niccol\`o Dalmasso,$^1$
  %\And
  Ann B. Lee,$^1$  
  %\And
  Rafael Izbicki,$^2$ %}
%\vspace{0.2cm}
%\aistatsauthor{
%\And
  Taylor Pospisil,$^{3}$   
%  \And
  Ilmun Kim,$^1$ 
%  \And
  Chieh-An Lin$^4$}

\vspace{0.75cm}

% \aistatsaddress{$^1$ Department of Statistics \& Data Science, 
%   Carnegie Mellon University University \\
%   $^2$ Department of Statistics,
%   Federal University of S\~ao Carlos \\
%   $^3$ Google LLC \\
%   $^4$ Institute for Astronomy,
%   University of Edinburgh } 
  
 ]

\begin{abstract}
 Complex phenomena in engineering and the sciences are often modeled with computationally intensive feed-forward simulations for which a tractable analytic likelihood does not exist. In these cases, it is sometimes necessary to estimate an approximate likelihood or fit a fast emulator model for efficient statistical inference; such surrogate models include Gaussian synthetic likelihoods and more recently neural density estimators such as autoregressive models and normalizing flows. 
 To date, however, there is no consistent way of quantifying the quality of such a fit. Here we propose a statistical framework that can distinguish any arbitrary misspecified model from  the target likelihood, and that in addition can identify with statistical confidence the regions of parameter as well as feature space where the fit is inadequate. Our validation method
 applies to settings where simulations are extremely costly and generated in batches or ``ensembles'' at fixed locations in parameter space. At the heart of our approach is a  two-sample test that quantifies the quality of the fit at fixed parameter values, and a global test that assesses goodness-of-fit across simulation parameters.
 While our general framework can incorporate any test statistic or distance metric, we specifically argue for a new two-sample test that can leverage any regression method to attain high power and provide diagnostics in complex data settings.
\end{abstract}

\section{Introduction}
\label{sec::introduction}
The likelihood function $\L(\x;\theta)$ links the unknown components $\theta$ of the data-generating mechanism with the observable data $\x$ and is a key component for performing statistical inference over parameters of interest.  For complex phenomena, there is often no tractable analytical form for the likelihood; many times such phenomena are instead studied using numerical simulators derived from the underlying physical or biological processes, which encode, e.g, complex observational effects, selection biases, etc.  In situations where
{the likelihood function cannot be easily evaluated}, but a stochastic numerical simulator {(which serves as the ground truth)} is available, approximate inference of parameters of interest is possible.
{Tools that explore feed-forward simulations {to infer $\theta$} without requiring explicit likelihoods are referred to as \emph{likelihood-free inference} (LFI) methods, of which Approximate Bayesian Computation (ABC) \citep{beaumont2002approximate,marin2012approximate} is the best known approach. Several variations of ABC methods exist and have resulted in many successful applications; see \cite{sisson2018handbook} for a review.}

However, there is a growing number of disciplines where
accurate analyses require highly realistic and computationally intensive simulations. 
{In such cases, it may not be feasible to repeatedly generate new simulations at different parameter settings as generally required by ABC methods.}
Instead, a common practice
is to 
run the simulator 
only for a few points in parameter space, 
in a format of batches or ensembles,  
 where an {\em ensemble} is a collection of multiple realizations (e.g., corresponding to different initial conditions) of the same physical model {(i.e. they all share the same $\theta$)}.  
 For example, modern climate and weather forecasting models (e.g., CESM \citep{hurrell2013community}) often 
incorporate complex representations of the atmosphere, ocean, land, ice, etc, on fine spatial and temporal resolutions across the entire world.
These models are commonly run as an ensemble of dynamical simulations with different initial conditions, where each simulation can take weeks to compile on supercomputer clusters (see \citep{baker2015new, kay2015CESMClimate} and references within). 
 Similarly, cosmological N-body simulations,
which compute gravity between particle pairs, are equally costly and often either created at a fixed cosmology (parameter value $\theta$) \citep{DES2016,KIDS2017},
or on a sparse grid of a few carefully chosen parameter values
 \citep{kacprzak2016cosmology,gupta2018non}.

Given the above scenario, a solution to make inference feasible is to  \add{replace the} computationally expensive simulator with a faster {\em emulator} model
that can speed up probabilistic modeling by several orders of magnitude. 
% \nic{11/27: If the goal is to infer parameters $\theta$ of interest, then these emulators forward-generate $x$ given $\theta$, hence providing an explicit approximation to the likelihood.\footnote{In this work we will use the terms {\em emulator} and {\em approximate likelihood} interchangeably to denote generative models that {directly}
%model  the relationship between observable data $\x$ and %parameters $\theta$.}}.
%\annlee{11/28, I further edited the text and footnote:} 
\add{If the goal is to infer parameters $\theta$ of interest, then these emulators often forward-generate $\x$ given $\theta$, thereby providing an explicit approximation to the likelihood.\footnote{In this work we will use the terms {\em emulator} and {\em approximate likelihood} interchangeably to denote generative models that {directly}
%model  
imply a relationship between observable data $\x$ and parameters $\theta$.}}
Some common \add{approximate or surrogate likelihood} models
 %\add{, which directly approximate the likelihood,}
%\footnote{In this work we will use the terms {\em emulator} and {\em approximate likelihood} interchangeably to denote generative models that {directly} model  the relationship between observable data $\x$ and parameters $\theta$.},} 
are Gaussian synthetic likelihoods 
\citep{wood2010statistical, price2018bayesian,Ong2018SyntheticLikelihoodGaussian}, 
 density ratio estimators \citep{izbicki2014,thomas2016,Dinev2018DIREDensityRatio}, and more recently neural density estimators (NDE), e.g., autoregressive models \citep{Uria2014RNADE-ar, Theis2015LSTMGenMod-ar, germain2015MADE-ar, Uria2016NADE-ar, Oord2016Wavenet-ar, Oord2016PixelCNN-ar, Oord2016PixelRNN-ar} and normalizing flows \citep{Dinh2014NICE-flow, Rezende2015VIN-flow, Dinh2016NVP-flow, Kingma2016IAF-flow, Papamakarios2017autoregressiveflow, Kingma2018Glow-flow, Chen2018NODE-flow, papamakarios2018sequential}. 
Other related works estimate the likelihood ratio \citep{Tong2013LikRatio, Cranmer2015LikRatio, Stoye2018LR, Brehmer2018LR} \add{and can also be used to emulate observable data as long as one can easily sample from the distribution in the denominator.}
%\ann{@Nic, check if that is true for the cited papers; I think it is if the denominator is the marginal distribution or a distribution that it is easy to generate data from.} \rafael{12/2 If it is true, I think it is a good edit}
\add{This paper is about how to assess any of the above LFI emulator models.}
(ABC and methods that \edit{directly estimate} the posterior instead of the likelihood, e.g., some Gaussian process emulators  \citep{Rasmussen2005GPM,wilkinson2014accelerating,meeds2014gps} and approximate posterior models \citep{papamakarios2016fast,lueckamnn2017Posterior, TALe2017ProbProgramming, izbicki2018ABCCDE, jarvenpaa2018Bolfi,greenberg2019automatic},  benefit from  %\remove{validation\rafael{12/1 We haven't discussed validation yet, so this sentence seems to be misplaced} techniques} \annlee{add ``more precise
%\rafael{'specific'? (I didn't understand the meaning of 'precise' here}
 \add{better assessment techniques} as well but we do not discuss them in this work.)

 %\ann{12/1 How about the following sentence?} %\annlee{``(We do not directly discuss ABC and methods that directly estimate the posterior instead of the likelihood, e.g., some Gaussian process emulators  \citep{Rasmussen2005GPM,wilkinson2014accelerating,meeds2014gps} and approximate posterior models \citep{papamakarios2016fast,lueckamnn2017Posterior, TALe2017ProbProgramming, izbicki2018ABCCDE, jarvenpaa2018Bolfi,greenberg2019automatic})''} 

Typically, machine learning-based LFI models are assessed   by computing built-in loss functions (e.g., Kullback-Leibler divergences in  
 %\remove{emulator networks}
  \add{autoregressive flows.})
 Such loss functions however only return a relative measure of performance rather than a goodness-of-fit to simulated data;  they do not  answer the question ``Should we keep searching for better estimates for this problem or is our fit good enough?'' (see Figure \ref{fig:synthetic-results}, left, for an example). Thus, an important  challenge is that of {\em validation}: determining whether an approximate likelihood or emulator model reproduces to the extent possible the targeted simulations in distribution. 
If the model is deemed inadequate, then the question of {\em diagnostics} becomes relevant. That is, pinpointing  ``how'' and ``where'' the emulator differs from the simulator in a potentially high-dimensional feature space across different parameters; {thereby providing valuable information for further improvements of the emulator, and insights on which simulations to run given a fixed budget.}  Up to now, popular approaches to simulation-based validation  \citep{cooks2006posteriorquantile,prangle2014diagnostic,talts2018validating} are valuable as consistency checks, but cannot always identify likelihood models that are clearly misspecified (see Section \ref{sec:examples_gof} for an example).
Furthermore, as these tools were originally designed for checking  
%\remove{Bayesian posterior models} \nic{11/27: "posterior approximations in Bayesian models" - R2 mentioned this was not completely clear},
 \add{posterior approximations in Bayesian models,}
they do not capture all aspects of the estimated likelihood, and therefore provide limited information on how to improve the estimates. 

In this paper, we propose general procedures for validating likelihood models; \add{after validation the approximate likelihood models can be used to compute approximate posteriors, or to construct confidence regions and hypothesis tests of parameters as outlined in Appendix~\ref{sec:approx_pval_conf_regions}.}
%\ann{instead of adding a clause after a semicolon, one could also alternatively put this text in parenthesis but I was worried that we would look to scatter-brained.}\rafael{I liked the semicolon version better as well}
  \add{Our validation} procedures are inspired by classical hypothesis testing, but generalize to complex data in an LFI setting,
and can identify statistically significant deviations from the simulated distribution. 
We use a new regression-based two-sample test \cite{kim2016techreport, kim2018regression}
to first compare the simulator and emulator models locally, i.e., at fixed parameters; 
these local tests are then aggregated into a ``global'' goodness-of-fit test that is statistically consistent  (see Theorem \ref{thm:main_control}). 
Our framework can adopt any machine-learning regression method to handle different structures in high-dimensional data. 
As Theorem \ref{Theorem: Global Regression Test} and Figure \ref{fig::synthetic_examples} show, this property translates to {\em high power} (for a fixed computational budget) under a variety of practical scenarios.

\textbf{Related Work.} 
Hypothesis testing has recently been used as a goodness-of-fit of generative adversarial networks (GANs \cite{Goodfellow2014GAN}); e.g., \citep{Jitkrittum2016Interpretable, Jitkrittum2017LinearTimeKernel, Jitkrittum2018InformativeFeatures} use two-sample tests to  detect features and feature space regions which discriminate between real and generated data. Although 
implicit (i.e. latent parameter) generative models 
are not our main focus, our local test has similar diagnostic capabilities, as shown in Appendix~\ref{sec:morphology}   and an earlier applied paper on galaxy morphology images \cite{Freeman2017LocalTestGalaxy}. There are also close connections between classification accuracy tests \citep{kim2016classification, LopezPaz2017C2ST} and our regression test. The main difference lies in the test statistic: classification accuracy tests are based on ``global'' error rates. Hence classifier tests can tell whether two distributions are different (i.e. they are two-sample tests) but these tests do not {\em per se} identify locally significant differences between two distributions with statistical confidence; for that one needs to consider the regression or class-conditional probabilities $\E(Y|\x)=\P(Y=1|\x)$ (where $Y$ here is the indicator function that $\x$ was generated by the emulator as opposed to the forward-simulator), which is the basis of our regression test statistic (Equation \ref{eq:regression_TS}).

\textbf{Novelty.}  
To date, there are no other validation technique in the LFI literature that 
can answer the following questions in a statistically rigorous way:
\vspace{-2.5mm}
\begin{enumerate}[(i)]
    \item {\bf if} one needs 
     %\remove{more simulations}
     to improve emulators for reliable inference
    %for improving emulators to obtain reliable inference 
    from observed data, i.e., whether the difference between the ``truth'' and the approximation learned with the existing train data is statistically significant; this question is answered by our %\remove{consistent} 
    global procedure (see Figure \ref{fig:synthetic-results}, left and Figure \ref{fig:linc-gof}, left);
    %\remove{and allows one to properly quantify cost-performance tradeoff,} \nic{11/27: removing the two above for R2/R3 comments. Adding an extra line below.}
    \vspace{-1.5mm} 
    \item {\bf where} in parameter space one, if needed, should propose the next batch of simulations; this question is answered by our local procedure (see Figure \ref{fig:synthetic-results}, right) and provides insights as to which simulations to run given a fixed budget; 
    and
    \vspace{-1.5mm}
    \item {\bf how} the distributions of emulated and high-resolution simulated data may differ in a potentially high-dimensional feature space; this question is answered by our regression test (see Figure \ref{fig:linc-gof},   right, and Appendix~\ref{sec:morphology})
    and offers valuable information as to what types of observations are under- or over-represented by the emulator and whether such differences are statistically significant. %\nic{11/27: Such insights can drive the decision to whether  employ a more powerful emulator model or to generate more simulations.}
    \add{Such insights can guide decisions as to whether it is necessary to improve the emulator model or generate more simulations.}
\end{enumerate}
\vspace{-2mm}

Moreover, we provide theoretical guarantees that ensure that %\remove{(a) the global test is able to tell if the estimated likelihood is wrong (i.e., no clearly misspecified models can pass the test; 
% Theorem \ref{thm:main_control}), and (b) the local test has high power as long as we have a good estimate of the regression function (Theorem \ref{Theorem: Global Regression Test})}. 
 %\nic{11/27: "(a) the local test has high power provided we have a good estimate of the regression function (Theorem \ref{Theorem: Global Regression Test}) and (b) the global test is able to tell if the estimated likelihood is wrong -- i.e., no clearly misspecified models can pass the test --  as long as the local test is statistically consistent (Theorem \ref{thm:main_control}). %\rafael{That is, the proposed tests have power against all alternatives (CITE Friedman, 2004)}"}
 %\annlee{(11/28 shortened further, no need to use the same terms ``clearly misspecified model'' or ``consistent'' in so many places):} 
 \add{(a) the local test has high power provided we have a good estimate of the regression function (Theorem \ref{Theorem: Global Regression Test}), and (b)  the global test has power against all alternatives \cite{friedman2004multivariate} 
  as long as the local test is statistically consistent (Theorem \ref{thm:main_control}).}
 %is able to tell if the estimated likelihood is wrong -- i.e., no clearly misspecified models can pass the test --  as long as the local test is statistically consistent (Theorem \ref{thm:main_control}). \rafael{That is, the proposed tests have power against all alternatives (CITE Friedman, 2004)}"}
\add{Our validation framework applies to any approximate likelihood model that can emulate $\x$ at parameter values $\theta$. The framework  applies to settings in which both simulations and emulations are cheap, the more challenging batch settings (described above), and settings with inexpensive emulations but very few simulations (see Appendix \ref{sec:test_MC_sampling}).}

\textbf{Organization.} The organization of the paper is as follows: In Section \ref{sec:validation} we describe our validation method, and provide theoretical guarantees 
as well as synthetic examples
that compare the performance of our goodness-of-fit test over some popular simulation-based calibration and distance-based tests. Then in Section \ref{sec::applications}, we show how our tools can be used to assess and diagnose models
for cosmological parameter inference.  Proofs of theorems and details on  
the high-dimensional sample comparison in feature space   
are provided in Appendix.

\textbf{Notation.} We indicate with $\mathcal{X}$ the feature space and with $\Theta$ the parameter settings where the 
%computationally expensive ensemble 
simulations from the ``true'' %model 
likelihood $\mathcal{L}(\x;\theta)$ are available. We denote the approximate likelihood from the emulator model by $\widehat{\mathcal{L}}(\x;\theta)$.
Both likelihood functions are normalized over 
$\mathcal{X}$; that is, $\int_\mathcal{X} \mathcal{L}(\x;\theta) d\x=  \int_\mathcal{X} \widehat{\mathcal{L}}(\x;\theta) d\x = 1$ for every $\theta \in \Theta$.

\section{Model Validation by Goodness-of-Fit Test} 
\label{sec:validation}

Our validation approach compares samples from the simulator with samples from the emulator, 
and can detect local discrepancies for a given parameter setting 
$\theta_0 \in \Theta$ as well as global discrepancies across parameter settings in $\Theta$. 
The validation procedure is as follows: For each $\theta_0 \in \Theta$, 
we first test the null hypothesis
$H_0:\widehat{\L}(\x;\theta_0)=\L(\x;\theta_0)$
for all $\x \in \mathcal{X}$. This {\em local test} (Algorithm \ref{alg:local}) compares output from the approximate likelihood/emulator model with a ``test sample'' from the simulator/true likelihood (the latter sample can be a held-out subset of a pre-generated ensemble at $\theta_0$ which has not been used to fit $\widehat{\L}(\x;\theta)$). A challenging problem is how to perform a two-sample test that is able to handle different types of data $\x$, and which in addition informs us on how  two samples differ in feature space $\mathcal{X}$; in Section \ref{sec::two-sample-regr} and Algorithm~\ref{alg:regression_test} we propose a new \emph{regression test}
that addresses both these questions. 
After the two-sample comparisons, we combine local assessments into a {\em global test} (Algorithm \ref{alg:global}) for checking if $\widehat{\mathcal{L}}(\x;\theta)=\mathcal{L}(\x;\theta)$ for all $\theta  \in \Theta$. 
The essence of the global test is to pool $p$-values which, under the null hypothesis, are uniform. Unlike many previous works  on pooling 
 $p$-values for multiple testing  
(e.g., \cite{lorenz2016does}), the $p$-values in Algorithm \ref{alg:global} are independent by construction.

The next section provides theoretical guarantees that the global test for our LFI setting is indeed consistent. These results apply for any sampling/weighting scheme $r(\theta)$ over $\Theta$ in Algorithm \ref{alg:global}, and for any consistent local test in Algorithm \ref{alg:local}. 

  \begin{algorithm} %[H]
  \caption{ \small Local Test for Fixed $\theta$}\label{alg:local}
  \algorithmicrequire \ {\small  
  parameter value $\theta_0$,
  two-sample testing procedure,
  number of draws from  the true model,
  $n_{{\rm sim},0}$
  and from the estimated model,
  $n_{{\rm sim},1}$}\\ 
  \algorithmicensure \ {\small $p$-value $p_{\theta_0}$
  for testing if
  $L(\x;\theta_0)=\widehat{L}(\x;\theta_0)$ for every  $\x \in \mathcal{X}$}
  \begin{algorithmic}[1]
     \STATE Sample $\mathcal{S}_0=\{\X_1^{\theta_0},\ldots,\X_{n_{\text{sim},0}}^{\theta_0}\}$ from $\mathcal{L}(\x;\theta_0)$.
        \STATE Sample $\mathcal{S}_1=\{\X_1^*,\ldots,\X_{n_{\text{sim},1}}^*\}$ from
 $\widehat{\L}(\x;\theta_0)$.
      \STATE
      Compute $p$-value
      $p_{\theta_0}$ for the comparison
      between $\mathcal{S}_0$ and
      $\mathcal{S}_1$.
     \STATE \textbf{return} {\small $p_{\theta_0}$}  
  \end{algorithmic}
  \end{algorithm}

  \begin{algorithm} %[H]
  \caption{ \small Global Test Across $\theta \in \Theta$ }\label{alg:global}
  \algorithmicrequire \ {\small  
  reference distribution $r(\theta)$, $B$,
  uniform testing procedure (e.g. Kolmogorov-Smirnoff, Cram\'er-von Mises)} \\
\algorithmicensure \ {\small $p$-value $p$ for testing if
  $L(\x;\theta)=\widehat{L}(\x;\theta)$ for every  $\x \in \mathcal{X}$ and $\theta \in \Theta$}
  \begin{algorithmic}[1]
  \FOR{$i \in \{1,\ldots,B\}$}
     \STATE sample $\theta_i \sim r(\theta)$
        \STATE compute $p_{\theta_i}$
        using Algorithm \ref{alg:local}
 \ENDFOR
      \STATE
      Compute $p$-value
      $p$    for testing if
      $\{p_{\theta_i}\}_{i=1}^B$  has a uniform distribution.
     \STATE \textbf{return} {\small $p$}  
  \end{algorithmic}
  \end{algorithm}

\subsection{Theoretical Guarantees for Global Test}
\label{sec::theory}
Here we provide sufficient assumptions for the global test to be  statistically consistent;  i.e., to be able to detect a misspecified distribution (as in Example 1) for large sample sizes.

\begin{Def}\label{def: global_test_def}
Define 
$\mathbb{D}_{B,n_{\text{sim}}}=\{p^{n_{\text{sim}}}_{\theta_1},\ldots,p^{n_{\text{sim}}}_{\theta_B}\},$ where
$p^{n_{\text{sim}}}_{\theta_1},\ldots,p^{n_{\text{sim}}}_{\theta_B}$
are the $p$-values obtained by Algorithm \ref{alg:local} using
$n_{\text{sim},1}=n_{\text{sim},2}=n_{\text{sim}}$,
and $\theta_1,\ldots,\theta_B \stackrel{\text{i.i.d.}}{\sim} r(\theta)$. Let $S(\mathbb{D}_{B,n_{\text{sim}}})$ be the test statistic for the global test.
Also, denote by
$S(\mathbb{U}_{B})$ the test statistic when
$\mathbb{U}_{B}=(U_1,\ldots,U_B)$,
with $U_1,\ldots,U_B \stackrel{\text{i.i.d.}}{\sim} U(0,1)$.
\end{Def}

\begin{Assumption} 
\label{assump:support}
Let 
$D=\left\{\theta:\mu_{\widehat{\L}(\cdot;\theta)}\neq \mu_{{\L}_\cdot(\theta)}\right\},$
where $\mu_{\widehat{\L}(\cdot;\theta)}$ ($\mu_{{\L}(\cdot;\theta)}$)
is the measure over $\mathcal{X}$ induced by ${\L}(\cdot;\theta)$ ($\widehat{\L}(\cdot;\theta)$).
Assume that $\mu_r(D)>0$, where $\mu_r$ is the measure  over $\Theta$ induced by $r(\theta)$.
\end{Assumption} 

\begin{Assumption}
\label{assump:convergece_local}
Assume that if 
$\theta_1 \in D$, then the local test is such that
$p^{n_{\text{sim}}}_{\theta_1} \xrightarrow[n_{\text{sim}}\longrightarrow \infty]{\P}0.$
Moreover, if 
$\theta_1 \notin D$, then the local test is such that
$p^{n_{\text{sim}}}_{\theta_1}\sim U(0,1).$
\end{Assumption}

\begin{Assumption} 
\label{assump:stat_zero}For every $0<\alpha<1$, the test statistic $S$ is such that
$F^{-1}_{S(\mathbb{U}_B)}(1-\alpha) \xrightarrow{B \longrightarrow \infty} 0.$
\end{Assumption} 

\begin{Assumption} 
\label{assump:uniformconsistent_stat}
%Assume that when Assumptions \ref{assump:support} and \ref{assump:convergece_local} hold,
Under Assumptions \ref{assump:support} and \ref{assump:convergece_local},
there exists $a>0$ such that
the test statistic $S$ satisfies
$S(\mathbb{D}_{B,n_{\text{sim}}})  \xrightarrow[B,n_{\text{sim}} \longrightarrow \infty]{\P} a.$
\end{Assumption} 

Assumption \ref{assump:support}
states that the set of parameter values
where the likelihood function is incorrectly estimated has positive mass under the reference distribution.
Assumption \ref{assump:convergece_local}
states  that the test chosen to perform the local comparisons is statistically consistent
and that its $p$-value has uniform distribution under the null hypothesis.
%Assumptions %\ref{assump:stat_zero} and
%\ref{assump:uniformconsistent_stat}
%state that the global test %statistic 
Assumptions \ref{assump:stat_zero} and
\ref{assump:uniformconsistent_stat} state that the test statistic for the global comparison in step 5 of Algorithm \ref{alg:global}
is statistically consistent, i.e.,  (i) it approaches zero under the null hypothesis when $B$ increases, and (ii) it converges to a positive number if the null hypothesis is false.  Under these four assumptions, we can guarantee statistical consistency.
%and the local test is consistent. 

\begin{thm}
\label{thm:main_control}
Let $\phi$ be an $\alpha$-level testing
procedure based on the global test statistic $S$. If the likelihood estimate
and the local and global test statistics are such that Assumptions
\ref{assump:support}--\ref{assump:uniformconsistent_stat} hold, then
$$\P\left(\phi_{S}(\mathbb{D}_{B,n_{\text{sim}}})=1\right)\xrightarrow{B,n_{\text{sim}} \longrightarrow \infty} 1$$ 
\end{thm}

\begin{Cor}
\label{cor:ksVon}
Under Assumptions \ref{assump:support} and
\ref{assump:convergece_local}, the global tests for comparing likelihood models based on Kolmogorov-Smirnoff and Cram\'er-von Mises  statistics are statistically consistent.
\end{Cor}

\subsection{Two-Sample Test via Regression}
\label{sec::two-sample-regr} 
Traditional approaches to comparing two distributions \citep{thas2010comparingdistr} 
\edit{often do not easily generalize} %are often not easily generalizable
to high-dimensional and non-Euclidean data. More recent non-parametric extensions (see \citep{hu2016review} for a review), e.g., maximum mean discrepancy (MMD) \cite{grettonMmdTest}, energy distance (ED) \cite{SzekelyEnergyTest}, divergence  \citep{sugiyama2011divergence,kanamori2012divergence}, mean embedding \citep{Chwialkowski2015MeTest, Jitkrittum2016Interpretable} and classification accuracy tests \cite{kim2016classification, LopezPaz2017C2ST} have shown to have power in high dimensions against some alternatives, specifically location and scale alternatives.

These methods, however, only provide a
binary answer of the form ``reject'' or ``fail to reject'' the null hypothesis. Here we propose a new regression-based approach to two-sample testing that can adapt to any structure in $\mathcal{X}$ where there is a suitable regression method; Theorem~\ref{Theorem: Global Regression Test}  relates the power of the test to the Mean Integrated Squared Error (MISE) of the regression. Moreover, the regression test can detect and describe  local differences (beyond the usual location and scale alternatives) in $\widehat{\L}(\x;\theta_0)$ and $\mathcal{L}(\x;\theta_0)$ in feature space  $\mathcal{X}$. We briefly describe the method below; see Appendix~\ref{sec:proof_two_sample_regr} and \citep{kim2018regression} for theoretical details, and see Sections \ref{sec:examples_gof} and \ref{sec::peak-count-data} for examples based on random forest regression. 

Let $P_0$ be the distribution over $\mathcal{X}$ induced by $\L(\x;\theta_0)$ and  let $P_1$ be the distribution over $\mathcal{X}$ induced by $\widehat{\L}(\x;\theta_0)$.
Assume that $P_0$ and $P_1$ have
density functions $f_0$ and $f_1$ relative a common dominating measure. By introducing  a random variable $Y \in \{0,1\}$ that indicates which distribution an observation belongs to, we can view $f_0$ and $f_1$ as conditional densities $f(\x|Y=0)$ and $f(\x|Y=1)$. 
The local null hypothesis is then equivalent to the hypothesis
$H_0: f_0(\x)= f_1(\x)$ for all $\x \in  \mathcal{X}_0:=\{\x \in \mathcal{X}: f(\x) > 0\}$, which in turn is  equivalent to
\vspace{-1mm}
$$H_0: ~ \P(Y=1 |\X=\x) = \P(Y=1), \ \ \text{for all } \x \in  \mathcal{X}_0.$$
We test $H_0$ against the alternative $H_1 : ~ \P(Y=1 |\X=\x) \neq  \P(Y=1), \ \ \text{for some~} \x \in  \mathcal{X}_0.$

By the above reformulation, we have converted the problem of two-sample testing to a {\em regression} problem. Depending on the choice of method for estimating the regression function $m(\x) = \P(Y=1 |\X=\x)$, we can adapt to nontraditional data settings involving mixed data types and various structures. 
More specifically, let $\widehat{m}(\x)$ be an estimate of $m(\x)$ based on the sample  $\{(\X_i,Y_i)\}_{i=1}^n$, and let $\widehat{\pi}_1 =  \frac{1}{n} \sum_{i=1}^n I(Y_i=1)$. We define our test statistic as 
\vspace{-1mm}
\begin{equation}\label{eq:regression_TS}
\widehat{\mathcal T} = \frac{1}{n} \sum_{i=1}^n \left( \widehat{m}(\X_i) - \widehat{\pi}_1 \right)^2.
\end{equation}
%\vspace{-1mm}
Note that the difference
$|\widehat{m}(\x) - \widehat{\pi}_1|$  {\em for each particular value of $\x \in \mathcal{X}$} also provides information on how well the emulator fits the simulator  locally {\em in feature space};  high values indicate a poor fit. To keep our framework as general as possible, we use a permutation procedure (Algorithm~\ref{alg:regression_test}) to  compute $p$-values; one could alternatively use a goodness-of-fit test via Monte Carlo sampling (see Remark~\ref{remark:MC_sampling} and Appendix~\ref{sec:test_MC_sampling}).

 Theorem~\ref{Theorem: Global Regression Test} shows that if $\widehat{m}$, the chosen regression estimator,
has a small MISE, the power of the test is large over a wide region of the alternative hypothesis. What this means in practice is that
 %\remove{ given a limited number of simulations,} 
we should choose a regression method that predicts the ``class membership'' $Y$ well.

\begin{algorithm}[!ht]
  \caption{ \small Two-Sample Regression Test via Permutations}\label{alg:regression_test}
  \algorithmicrequire \ {\small two i.i.d. samples $\mathcal{S}_0$ and $\mathcal{S}_1$ from distributions with resp. densities $f_0$ and  $f_1$; number of permutations $M$; a regression method $\widehat{m}$}\\ 
   \algorithmicensure \ {\small $p$-value for testing if $f_0(\x)=f_1(\x)$ for every $\x \in \mathcal{X}$}
    \begin{algorithmic}[1]
    \STATE Define an augmented sample   $\{\X_i, Y_i\}_{i=1}^n$, where $\{\X_i\}_{i=1}^n \!\!= \!\! \mathcal{S}_0  \cup \mathcal{S}_1$, and $Y_i=I(\X_i \in  \mathcal{S}_1)$.
     \STATE Calculate the test statistic $\widehat{\mathcal T}$ in Equation~\ref{eq:regression_TS}.
     \STATE  Randomly permute $\{Y_1,\ldots,Y_n\}$. Refit $\widehat{m}$ and calculate the test statistic on the permuted data.
     \STATE Repeat the previous step $M$ times to obtain $\big\{ \widehat{\mathcal T}^{(1)},  \ldots, \widehat{\mathcal T}^{(M)}\big\}$.
     \STATE Approximate the permutation $p$-value by
    		$p = \frac{1}{M+1} \left( 1 + \sum_{m=1}^M I(\widehat{\mathcal T}^{(m)} > \widehat{\mathcal T}  ) \right).$
     \STATE \textbf{return} $p$  
  \end{algorithmic}
\end{algorithm}

\begin{thm} \label{Theorem: Global Regression Test}
Suppose that the regression estimator  $\widehat{m}(\x)$ is  a linear smoother satisfying
$\sup_{m \in \mathcal{M}} \E \int_{\mathcal X} \left( \widehat{m}(\x) - m(\x) \right)^2 dP_X(\x) \leq C_0 \delta_n,$
where $C_0$ is a positive constant, $\delta_n = o(1)$, $\delta_n \geq n^{-1}$, and $\mathcal{M}$ is a class of regressions $m(\x)$ containing constant functions.
	 Let $t^*_\alpha$ be the upper $\alpha$ quantile of the permutation distribution of the test statistic $\widehat{\mathcal{T}}^\prime $ on validation data from sample splitting.\footnote{The proof assumes sample splitting where (for simplicity) half of the data is used to estimate the regression function and the other half is used to estimate the test statistic; {i.e., $\widehat{\mathcal T}' = 2n^{-1} \sum_{i={n/2}}^{n} \left( \widehat{m}(\X_i) - \widehat{\pi}_1 \right)^2
,$ where $\widehat{m}$ is estimated using $(\X_1,Y_1),\ldots,(\X_{n/2},Y_{n/2})$}.} 
	 Then for any $\alpha,\beta \in (0,1/2)$ and $n$ sufficiently large, there exists a universal constant $C_1$ such that %\annlee{can we reduce the spacing here?}
	 \vspace{-2mm}
	\begin{align*}
	&\text{Type I error: } \P_0 \left( \widehat{\mathcal{T}}^\prime  \geq t^\ast_\alpha \right) \leq \alpha, \\
	&\text{Type II error: } \sup_{m \in \mathcal{M}(C_1 \delta_n) } \P_1 \left( \widehat{\mathcal{T}}^\prime  < t^*_\alpha \right) \leq \beta
	\end{align*}
	\vspace{-2mm}
 against the class of alternatives $\mathcal{M}(C_1  \delta_n):=
\Big\{ m \in \mathcal{M}: \int_{\mathcal{X}} \left( m(\x) - \pi_1 \right)^2 dP_X(\x) \geq C_1 \delta_n \Big\}.$
\end{thm}

\begin{remark} [Goodness-of-fit via Monte Carlo sampling]\label{remark:MC_sampling}
 If the total number of test simulations from $\L(\x;\theta_0)$ is small, say \edit{$n_{\text sim} \lesssim 30$}, but the cost of drawing samples from the emulator model $\widehat{\L}(\x;\theta_0)$ is negligible, then one can instead of the two-sample permutation test in Algorithm~\ref{alg:regression_test} perform a goodness-of-fit test via repeated Monte Carlo sampling from the emulator (see Algorithm~\ref{alg:monte_carlo_regression_test} in Appendix~\ref{sec:test_MC_sampling} for details). In the goodness-of-fit formulation, one regards the emulator fit $\widehat{\L}(\x;\theta_0)$ as a reference distribution. To test that the simulated sample from $\L$, $S=\{ \X_1, \ldots, \X_{n_{\text sim}}  \}$,  is consistent with the reference distribution $\widehat{\L}$, one  
draws several independent large Monte Carlo (MC) samples $\{S^{(m)}\}_{m=1}^{M}$ from $\widehat{\L}(\x;\theta_0)$
 to produce a set of values $\{\widehat{T}^{(m)}\}_{m=1}^{M}$ that are used as a null distribution
 for $H_0: \L(\x;\theta_0)=\widehat{\L}(\x;\theta_0)$. To cite Friedman \citep[Section IV]{friedman2004multivariate}, such an approach has  ``the potential for increased power [compared to the two-sample permutation test] at the expense of having to generate many Monte Carlo samples, instead of just one''. Corollary \ref{cor:mc_sampling} states that our main result (Theorem \ref{Theorem: Global Regression Test}) still holds for the alternative goodness-of-fit test.
\end{remark}

\subsection{Examples}
\label{sec:examples_gof}

\begin{table*}[!ht]
\centering
\begin{tabular}{|c||c|c|c|}
\hline
\textit{Example 2 Settings} & \textit{True Likelihood $\mathcal{L}(\x;\theta)$} & \textit{Approx. Likelihood $\widehat{\mathcal{L}}(\x;\theta)$} & \textit{Parameter Region $\Theta$} \\ \hline
(a) Bernoulli            &   $\text{Bern}(x_1; \theta) \prod_{d=2}^{D} \mathcal{N}(x_d; \theta, 1)$  &  $\prod_{d=1}^{D} \mathcal{N}(x_d; \theta, 1)$  & $(0,1)$ \\[2pt] \hline
(b) Scaling             &   $\mathcal{N}(x_1; 0, \theta) \prod_{d=2}^{D} \mathcal{N}(x_d; 0, 1)$ & $\prod_{d=1}^{D} \mathcal{N}(x_d; 0, 1)$ & $(0,1)$ \\[2pt] \hline
 (c) Mixture of & $f_m(x_1; \theta, 1) \prod_{d=2}^{D} \mathcal{N}(x_d; 0, 1),$ where  & $\prod_{d=1}^{D} \mathcal{N}(x_d; 0, 1)$ &  $(-5,5)$ \\[2pt]
Gaussians & $f_m(\theta, 1) = 1/2 \mathcal{N}(-\theta, 1) + 1/2 \mathcal{N}(\theta, 1)$ & & \\ \hline
\end{tabular}
\caption{\small{The three toy settings in Example 2. In each setting, the true and approximate likelihood differ only in the first dimension, $x_1$. 
({$\mathcal{N}(x; \mu, \sigma^2)$} is a 1D Gaussian with mean $\mu$ and variance $\sigma^2$; $\text{Bern}(x; \theta)$ is a Bernoulli with parameter $\theta$.)}}\label{tab: example_2}
\end{table*}

We next use two synthetic examples to illustrate the advantages of our global and local tests to state-of-the-art validation techniques, in terms of consistency and higher power respectively.

{\bf Example 1 (Consistency of Global Test).} 
One key property of our global goodness-of-fit test is that 
it can detect any misspecified approximation of the likelihood function 
(Theorem \ref{thm:main_control}). 
Diagnostic tools like the Posterior Quantiles (PQ) technique \citep{cooks2006posteriorquantile} 
and Simulation-Based Calibration (SBC) \citep{talts2018validating} are  often used
to validate approximate likelihood models (see, e.g., \citep{papamakarios2018sequential}) by checking whether a histogram of respective statistics (posterior quantiles and ranks) is close to uniform. 
However, these tests are sometimes not able to tell the difference between the true model and 
a clearly misspecified model as illustrated by the following toy example, where
$\theta_i \sim \mbox{Gamma}(1,1)$, $i=1,\ldots,500$, and $\x=\edit{x}_1,\ldots,\edit{x}_{1000}|\theta_i \sim \mbox{Beta}(\theta_i, \theta_i)$. 

%\annlee{ended up keeping this paragraph; let me know if it's too much} 
\comment{The PQ test is based on the fact that, given  a sample $\tilde{\theta}$ from the prior distribution,   %\begin{equation*}
the posterior quantile
    $q(\tilde{\theta}) = \int f(\theta|\x) \mathbb{I}(\theta < \tilde{\theta}) d\theta$
%\end{equation*}
is uniformly distributed. Similarly, the SBC test relies on the fact %as well and the property 
that, given any ranking function $g(\theta)$ and a posterior sample $\{\theta_1,\ldots,\theta_L\}$,  the rank 
%\begin{equation*}
$r\left(g(\theta_1), ..., g(\theta_L), g(\tilde{\theta})\right) = \sum_{l = 1}^L \mathbb{I}(g(\theta_l) < g(\tilde{\theta}))$ is  uniformly distributed. Both PQ and SBC assess goodness-of-fit by checking if a histogram of respective statistics (posterior quantiles and ranks) is close to uniform.}
 
\begin{figure}[!ht]
\includegraphics[width=0.49\textwidth]{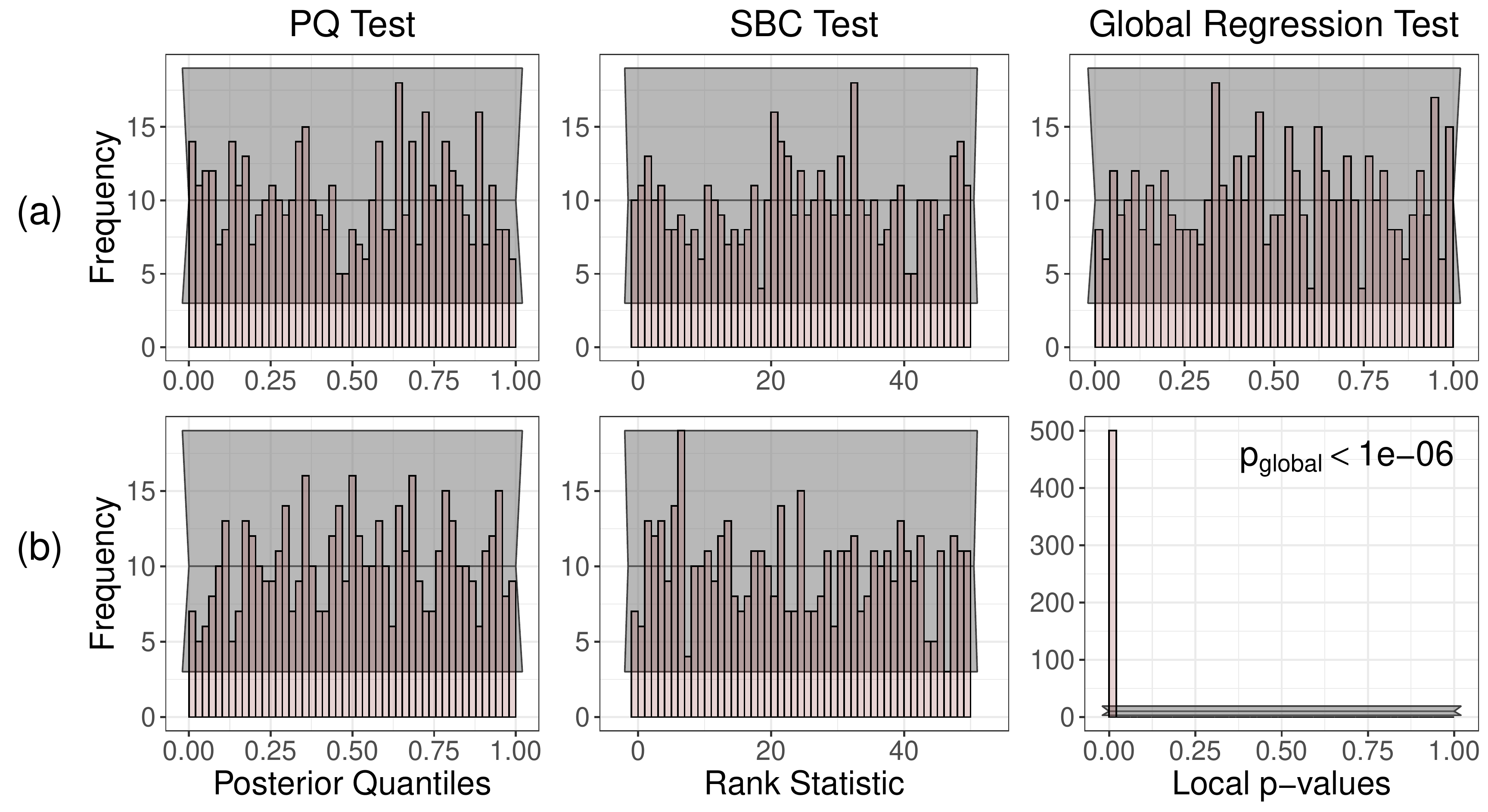}
\caption{\small Distribution of posterior quantiles, rank statistics and $p$-values for PQ, SBC and our global regression test, respectively, for (a) the true model in Example 1, and (b)  a clearly misspecified model. % where $\widehat{\mathcal{L}}_\x(\theta) \propto 1$. 
Only the global regression test correctly rejects the latter (bottom right plot). (The grey ribbon represents the 99\% confidence interval for the test of uniformity used for PQ and SBC.)
} \label{fig: global_consistency_example} 
\end{figure}

Figure \ref{fig: global_consistency_example} shows the distribution of the statistics computed for PQ and SBC (along with confidence regions that describe what one would expect under uniformity) and the distribution of our local $p$-values 
(recall that our global test
 is based on
 testing whether the local $p$-values are uniformly distributed) for two different cases: 
In the top row,  we consider a case where 
$\widehat{\mathcal{L}}(\x;\theta) =\mathcal{L}(\x;\theta)$.
All tests pass the model, as they should. 
In the bottom row, 
we consider a case where $\widehat{\mathcal{L}}(\x;\theta) \propto 1$, 
a poor approximation of the likelihood function (see Appendix~A for examples). 
Our global regression test, which is based on uniformity of the local $p$-values, clearly rejects this model. PQ and SBC, on the other hand,
cannot
  distinguish between the true likelihood and the misspecified model as these by construction have the same marginal distribution over $\theta$ in this toy example. 
Similar results \citep{desc_photoz} have been found for diagnostic tests of conditional density estimates when using quantities related to PQ and SBC (such as, PIT scores and QQ plots).

\begin{figure}[!ht]
\centering
\includegraphics[width=0.49\textwidth]{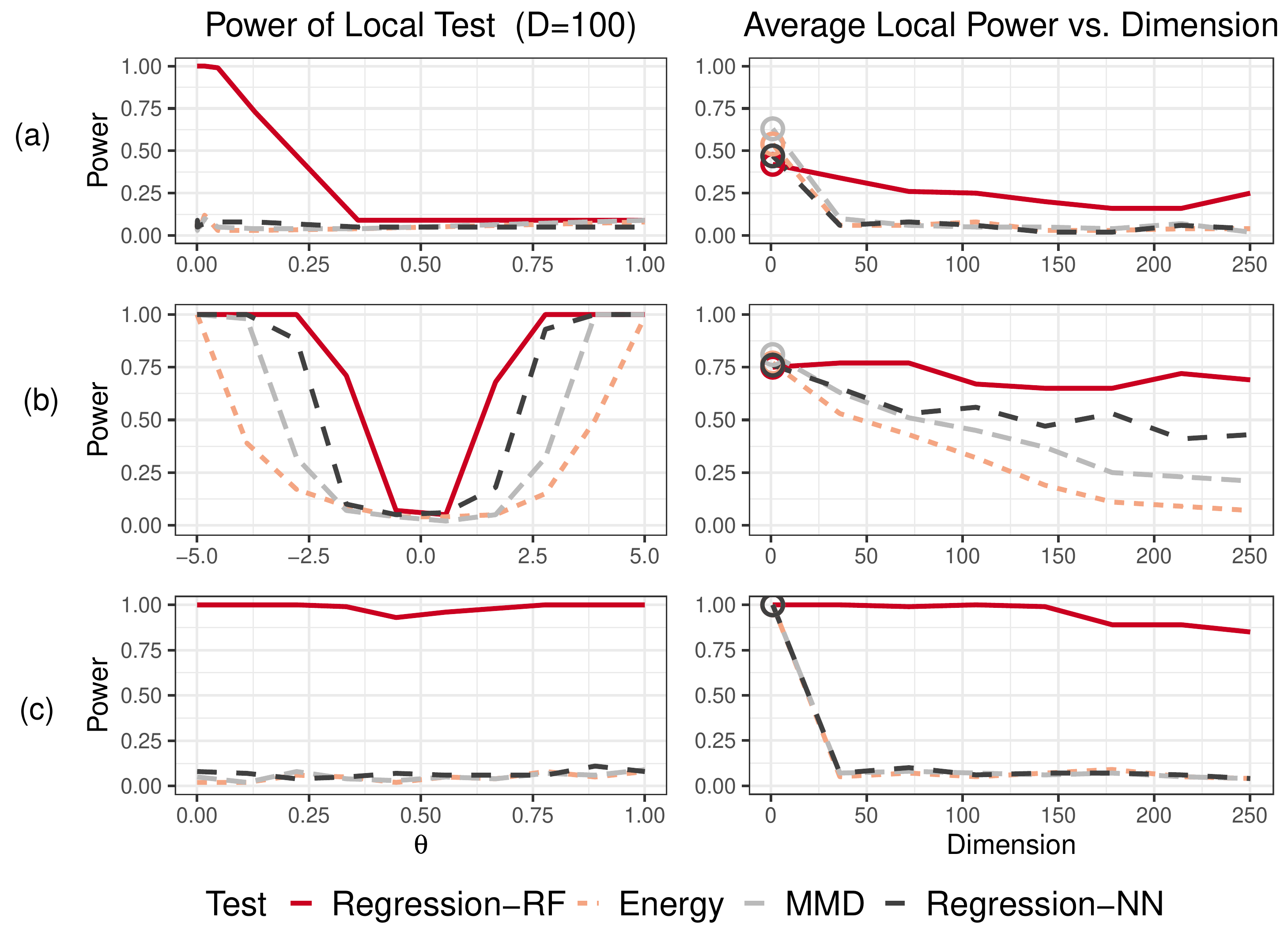}
\caption{\small 
 Local test power shown in the left column as a function of  $\theta$ at $D$=100, and shown in the right column as a function of the dimension $D$ (averaged over $\theta$) for the (a) Bernoulli, (b) Scaling, and (c) Mixture of Gaussians case.
Note that distance-based tests are more powerful at $D=1$ (highlighted with circles in the right column), but their power is severely affected with increasing dimension. Our {RF regression test achieves}  higher power for large $D$ by leveraging
the advantages of random forest regression in high-dimensional settings with sparse structure.
} \label{fig::synthetic_examples} 
\end{figure}

{\bf Example 2 (Power of Local Test).} 
The power of our goodness-of-fit test will much depend  on how we compare samples at fixed $\theta_0  \in \Theta$; that is, on how we test the local
null hypothesis, $H_0:\widehat{\L}(\x;\theta_0)=\L(\x;\theta_0)$ for every $\x \in \mathcal{X}$. An advantage of the regression approach (Algorithm \ref{alg:regression_test})
is that we can use any regression technique that  efficiently explores the structure of the data at hand; the practical implications of Theorem \ref{Theorem: Global Regression Test} is that one should choose the regression method with the smallest MISE (a quantity that can be estimated from data) to attain a higher test power (an unknown quantity).
We illustrate {these ideas} with a synthetic example where
$\x \in \mathbb{R}^D$, where $D$ could be large. We consider three  toy settings where the approximate likelihood  and the true likelihood  only differ in the first dimension --- that is, we test against a sparse alternative; see Table \ref{tab: example_2} for details.

\begin{figure*}[!ht]
\centering
\includegraphics[width=0.45\textwidth]{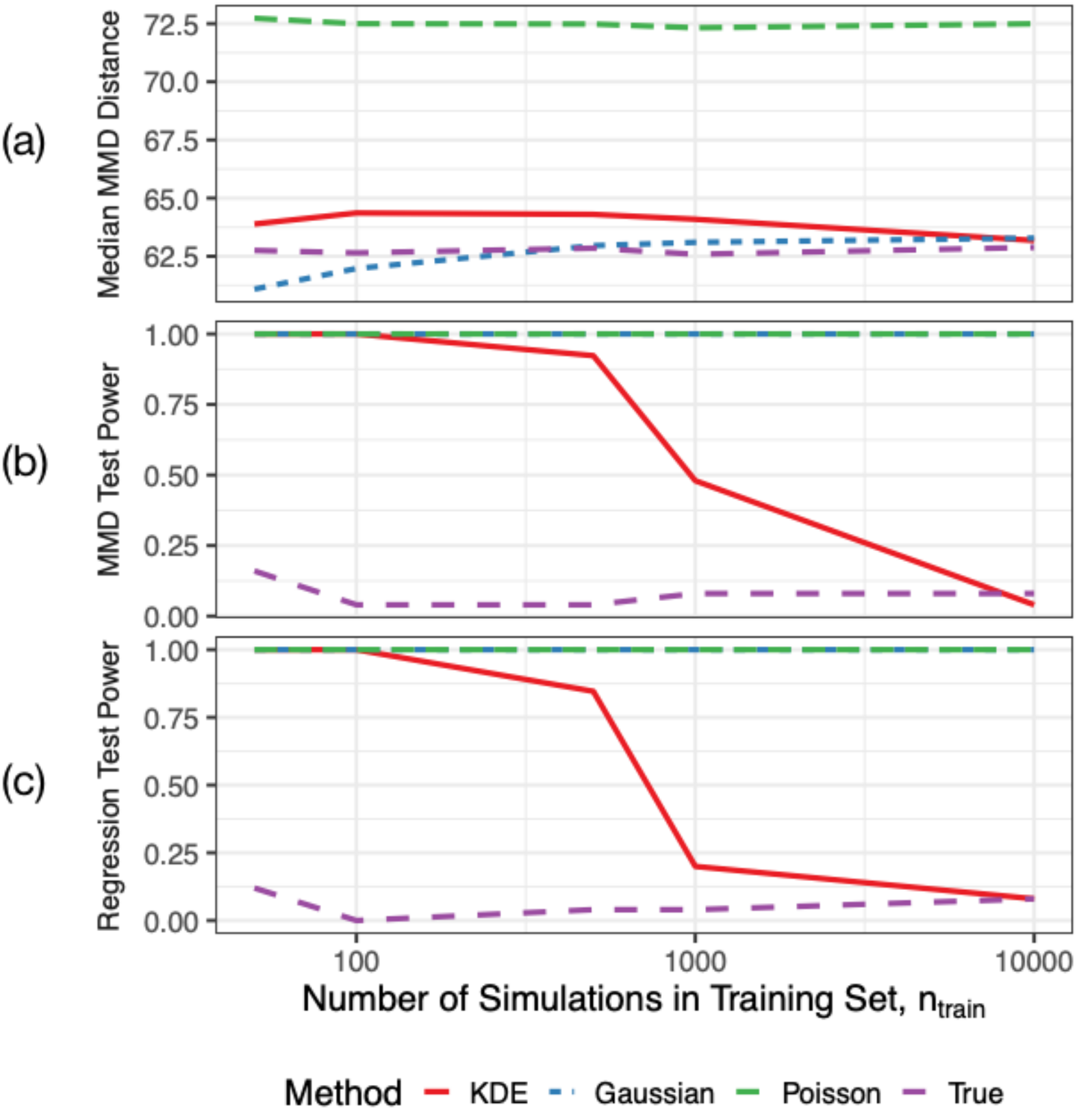}
\includegraphics[width=0.48\textwidth]{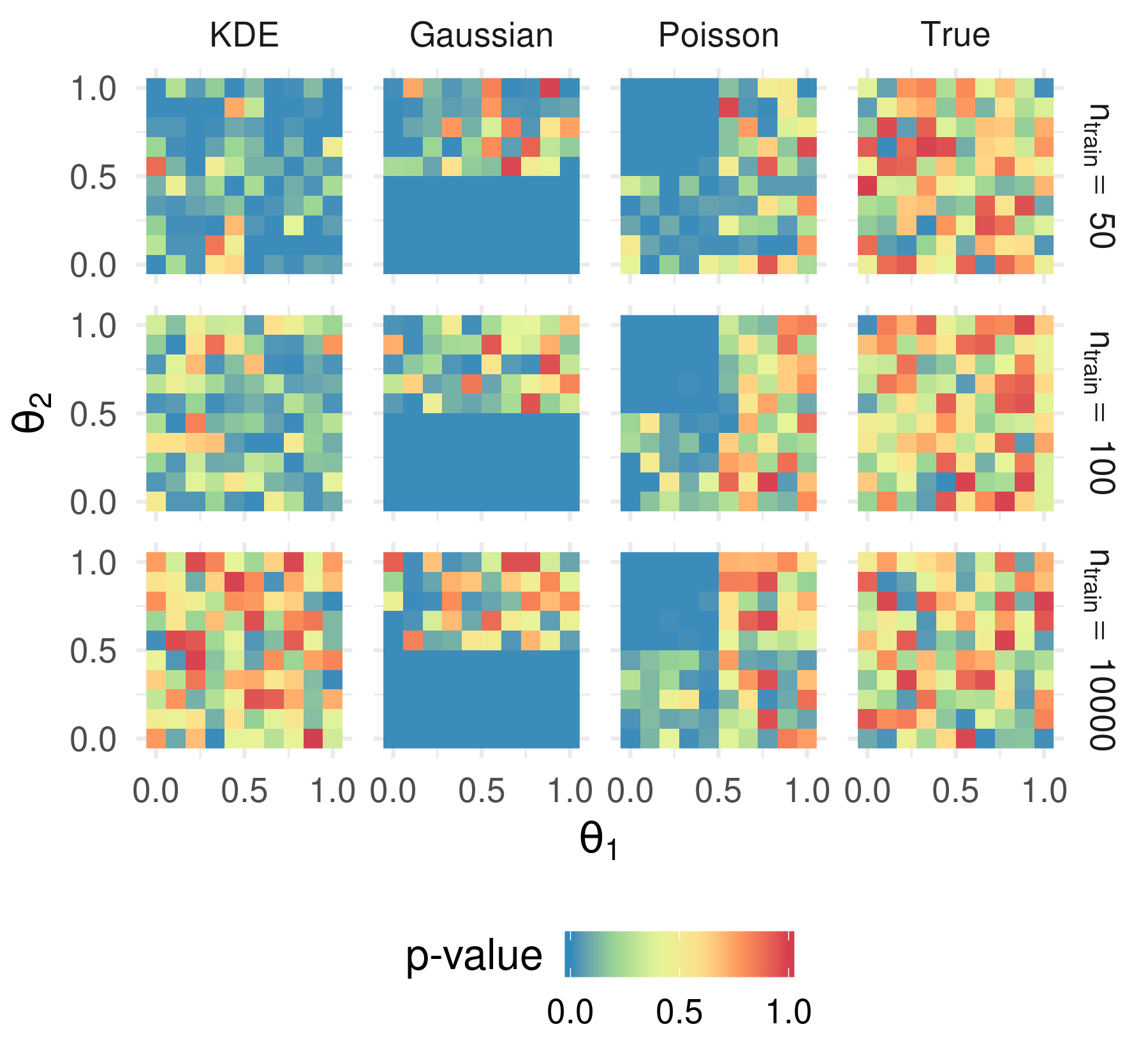}
\caption{\small {\em Left panel}: Median MMD distance and power of global goodness-of-fit tests (100 trials with $\alpha = 0.05$ and $n_{\text{sim}}=200$) for the synthetic example.  
(a) The median MMD over all trials is not informative as it  
 does not vary with  $n_{\text{train}}$. However, global tests based on (b) the MMD distance and (c) the regression are able to capture that KDE improves as $n_{\text{train}}$ increases 
 (the power decreases with $n_{\text{train}}$), while the parametric models do not. 
 {\em Right panel}: Local test $p$-values for regression test by model and number of training simulations. We can identify regions where models fit poorly: e.g., the Gaussian model fits poorly for bottom half of $\theta$-space as low counts cannot be adequately approximated as Gaussian.}
\label{fig:synthetic-results}
\end{figure*}

For each $\theta \in \Theta$, we compute a local $p$-value by comparing samples of size $n=100$ from  $\mathcal{L}(\x; \theta)$ and  $\widehat{\mathcal{L}}(\x; \theta)$, respectively (Algorithm \ref{alg:local}). This procedure is repeated $100$ times to estimate the  power function.
We apply the local test for three different test statistics; namely: (i) the test statistic in Equation~\ref{eq:regression_TS} using random forest  {(RF)} or nearest neighbor regression {(NN)},
   (ii) the MMD test statistic \citep[Eq. 5]{grettonMmdTest} with a Gaussian kernel, and  
    (iii) the energy test statistic \citep[Eq. 5]{BaringhausEnergyTest, SzekelyEnergyTest} using the Euclidean norm. 
Figure \ref{fig::synthetic_examples} shows how the power function varies with $\theta$ at dimension $D=100$ (left column) and how
the power, averaged over $\theta$, varies with $D$ (right column) for each setting.
When $D=1$ (highlighted with circles in the right column) distance-based tests  {based on RF} yield higher power, but their performance quickly degrades with increasing $D$. On the other hand,  our RF regression test is able to achieve higher power in high-dimensional settings by leveraging some advantages of random forest regression {(as shown by the red curves)}; such as, the ability to select features, and the ability to tell discrete versus continuous distributions apart. 
For instance,  {in the Bernoulli case (top row, a)}  our regression test 
has higher power for small values of $\theta$, 
which is when the distribution of the first coordinate is almost degenerate at 0.

\vspace{-2mm}
\section{Applications}
\label{sec::applications}

In this section we focus on \emph{validating} approximate likelihood models for 
 {cosmological parameter inference} with weak lensing peak counts. Weak lensing (WL) is a gravitational deflection effect of light by matter in the Universe that causes distortion in projected images of distant galaxies along lines of sight. We can use this effect to
 %\remove{constrain} \nic{11/27: R2 brought up that "constrain" is not completely clear. Could we substitute it with something like "estimate" or "learn" or "estimate the distribution of"? I am under the impression that "constrain" refers to characterize the potential values of a parameter when some/all the other model parameters are fixed.}\rafael{I like 'estimate'} 
  \add{estimate} parameters of the $\Lambda CDM$ cosmological model, the most well-supported model within Big Bang cosmology. In particular we can \add{estimate} the dark matter density $\Omega_{m}$ and 
  its clumpiness $\sigma_{8}$ 
 through {\em peak counts}: the number of local maxima in the WL   convergence map (a 2D image) binned by the value of the peak \cite{Dietrich2010}. 
In our data example, we use the \texttt{CAMELUS} simulator \cite{lin2015new} to generate peaks. 
 
  In Section \ref{sec::synthetic-example}, we showcase our approach on a synthetic example with known likelihood and properties similar to those of peak counts. To estimate the likelihood  we use two parametric models, a Gaussian and a Poisson model, as well as a non-parametric kernel density estimator (KDE) with the bandwidth estimated coordinate-wise according to \citep{Wand1994kdebandwidth} 
and discretized to reflect the integer-valued data. (The Gaussian model with a fixed covariance and varying mean is the  current state-of-the-art in cosmological parameter inference \cite{kacprzak2016cosmology}.) In Section \ref{sec::peak-count-data}, we provide results and insights 
with data obtained from the \texttt{CAMELUS} simulator, comparing the   %\remove{Gaussian and Poisson}
two parametric models with a conditional masked autoregressive flow (MAF; %\remove{\citep{Papamakarios2017autoregressiveflow}}
\citep{Papamakarios2017autoregressiveflow, papamakarios2018sequential}), again discretized to  reflect the integer-valued data.

\begin{figure*}[!ht]
\centering
\includegraphics[width=0.47\textwidth]{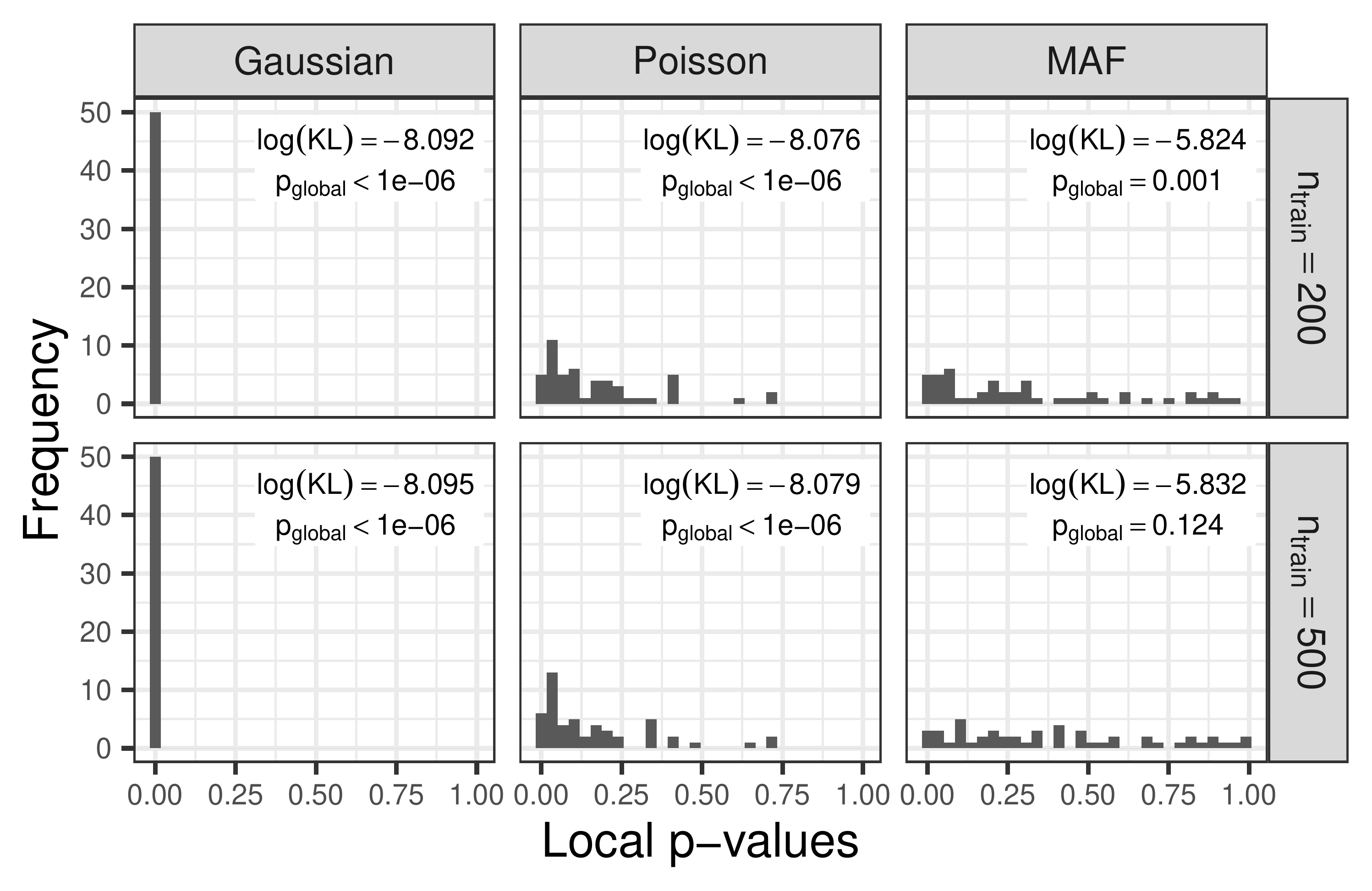}
\includegraphics[width=0.515\textwidth]{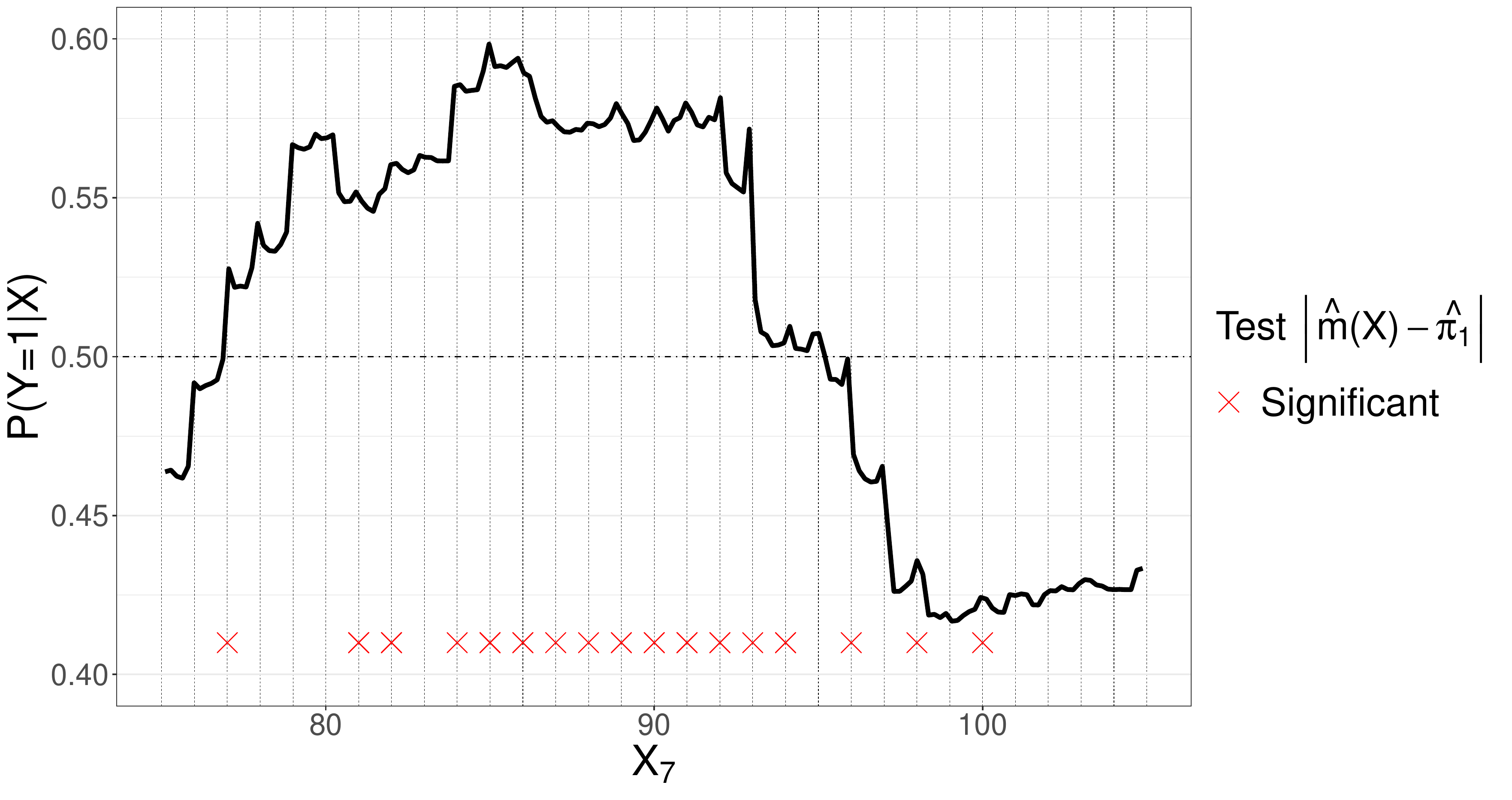}
\caption{\small {\em Left panel}: Local goodness-of-fit for peak-count data with $n_{\text{train}} = 200$ (top row) and $n_{\text{train}} = 500$ (bottom row). Although the Gaussian model is achieving the lowest KL divergence, the estimates are rejected at almost all $\theta$; Poisson and MAF perform better (more uniform-looking distributions of local $p$-values) but only MAF passes the global test at $n_{\text{train}} = 500$.
{\em Right panel}: 
 Partial dependence plot for variable $x_{7}$ (low count bin) for Gaussian model at $n_{\text{train}} = 200$. 
  The red crosses on the x-axis represent the locations where the difference $|\widehat{m}(\x) - \widehat{\pi}_1|$ is statistically significant according to a joint analysis in 7 dimensions; these locations coincide with integer values of $x_7$ and indicate that the regression test is distinguishing between the discrete true distribution of counts and the fitted continuous Gaussian distribution.
 }
\label{fig:linc-gof}
\end{figure*}

\subsection{Synthetic Example}
\label{sec::synthetic-example}

Peak count data possess two important properties: 
(1) data are discrete and 
(2) counts in different bins are correlated to each other. 
The first property implies that at high bin counts the data are approximately normally distributed, 
but for bins with low counts this approximation breaks down. 
The latter property introduces difficulties in modeling number counts as independent Poisson variables.
We mimic these two
properties by drawing 
$X_{1}, X_{2} \overset{\mbox{\tiny indep}}{\sim} \operatorname{Poisson}(\lambda)$, {where $\lambda$ depends on the parameter $\theta = [\theta_1, \theta_2] \in \mathbb{R}^2$}. 
When $\theta_{1} < 0.5$, we set $\lambda = 1$, otherwise $\lambda = 10^4$ 
which makes the normal approximation appropriate due to the Central Limit Theorem. 
When $\theta_{2} < 0.5$ we add the requirement that $X_{1} \le X_{2}$ (which breaks independence).
Our first experiment is to use our global test to assess likelihood models that are fit with different number of simulations ($n_{train} = [50, 100, 10000]$) while holding the size of the test samples fixed ($n_{\text{sim}} = 200$).
For the likelihood models mentioned above -- KDE, Gaussian and Poisson -- we implement Algorithm \ref{alg:global} with a uniform reference distribution
over a grid of 100 $\theta$-values evenly spaced in $[0,1]\times[0,1]$.
 For illustrative purposes, we conduct 100 trials resampling the entire dataset
to estimate the power of the test; that is, the probability that the global test procedure  rejects a likelihood model.
The fit of the likelihood models are assessed using three  criteria: (i) the median (over 100 trials) MMD distance between the two samples, (ii) the power of a global test based on the same MMD distances, and (iii) the power of a global regression test with random forest. It is common practice to compare emulator models (see, e.g., \cite{papamakarios2018sequential,greenberg2019automatic}) by computing distances such as MMD, which we here refer to as raw test statistics. Figure \ref{fig:synthetic-results}, left, shows that the ``Median MMD Distance'' (top, a) is not particularly informative in this example. On the other hand, the ``MMD Test Power'' (center, b) and the ``Regression Test Power'' (bottom, c)  tell  us that both the Poisson and Gaussian models are misspecified; these models are rejected regardless of $n_{\text{train}}$, whereas the KDE model slowly improves with the number of simulations until ultimately achieving a power similar to the true distribution. These results illustrate that the local and global $p$-values can be more informative than the test statistics themselves.

To better understand why the Gaussian and Poisson models fit poorly
 %achieve poor fits
  we can turn to the local information we calculated for each $\theta$ via Algorithm \ref{alg:local} ($n_{sim}=200$).
The data-generating process in our synthetic example induces four quadrants with different
behaviors. Figure \ref{fig:synthetic-results}, right, showcases the utility of the
local test: it pinpoints \emph{where} in the parameter space the model fits are insufficient. 
More specifically: for the Poisson fits, the $p$-values in the left $(\theta_{1} < 0.5)$ region are very small as are the $p$-values for the lower $(\theta_{2} < 0.5)$ region for the Gaussian fits. This is due to
the independence and Gaussian assumptions, respectively, breaking down in these two regions.
In addition, the KDE model improves as the number of simulations used to train the models increases, starting at poor fits with low $p$-values at
$n_{\text{train}}=50$ and eventually
achieving $p$-values drawn from the uniform distribution for large values of $n_{\text{train}}$. Our global test  makes this observation rigorous.

\subsection{Peak Count Data Example}
\label{sec::peak-count-data}

For WL peaks, we consider a 2D parameter space over $\theta = (\Omega_{m}, \sigma_8)$
and design a grid of 50 different cosmologies $\theta$ around a fiducial (probable) cosmology $\theta_0$ (see Appendix~D).
For each $\theta$-value, we
  simulate a batch of WL maps ($n_{\text{train}} = 200$,  $n_{\text{sim}} = 200$). The peak count data (i.e. histogram of peak intensities in each map) is a vector $\x \in \mathbb{N}^D$
where $D=7$ is the number of bins.
 We compare three approximate likelihood models: Gaussian, Poisson and conditional MAF.
To assess models, we first compute the Kullback-Leibler (KL) divergence loss 
for the $n_{sim}=200$ test simulations at each $\theta$.
 According to the KL loss, the Gaussian model performs best; however, these are only relative comparisons. 
We now use a RF regression test to find out whether the Gaussian model actually fits the simulated data well.  
As indicated in Figure \ref{fig:linc-gof} (left, top row), the local tests for the Gaussian model reject the
null hypothesis $\widehat{\L}(\x;\theta)=\L(\x;\theta)$ at almost every $\theta$; thus the global hypothesis is also rejected. The Poisson and MAF models are rejected by the global test as well but have a more uniform-looking distribution of local $p$-values.  Now if we increase the the number of train simulations to $n_{\text{train}}=500$ (while holding $n_{\text{sim}} = 200$ fixed), the fitted MAF model passes the global test whereas the Gaussian and Poisson models still do not as indicated by the bottom row (these qualitative results stay the same for $n_{\text{train}}=5000$).

Finally, our local regression tests can provide insights into \emph{how} the two distributions $\widehat{\mathcal{L}}(\x;\theta)$ and  $\mathcal{L}(\x;\theta)$ differ in feature space $\mathcal{X}$; more specifically, by evaluating how the estimate of the regression function $\widehat{m}(\x)$ in Equation \ref{eq:regression_TS} varies with $\x$ for a fixed $\theta$ (a significant difference $|\widehat{m}(\x)-\widehat{\pi}_1|$ is an indication that the model is not well estimated at that location in feature space)
We illustrate such an analysis for our fitted Gaussian model for  $n_{\text{train}} = 200$ and $\theta=\theta_0$. According to the RF regression used to construct our test statistic, the most influential variables correspond to bins with low counts. In Figure \ref{fig:linc-gof}, right, we visualize the fit on such a bin (variable $x_7$) by a partial dependence plot (which shows the marginal effect of this variable on $\widehat{m}(\x)$ \cite{friedman2001greedy}). On the $x$-axis, we mark the locations where the difference $|\widehat{m}(\x)-\widehat{\pi}_1|$ is statistically significant according to a joint analysis in 7 dimensions (see Algorithm~\ref{alg:local_test_difference} in Appendix~\ref{sec:morphology} for details). These locations coincide with integer values of $x_7$, showing that the regression test is distinguishing between the discrete true distribution for bin counts and the fitted continuous Gaussian distribution (these results also explain why the Poisson model may fare better).  In Appendix~\ref{sec:morphology}, we provide a detailed analysis of how one can identify and  visualize areas of significant differences in multivariate distributions for galaxy morphology images.

\section{Final Remarks}
 We have developed validation methods of approximate emulator models 
    that are able to identify a misspecified model and give insights on how to improve such a model; more specifically, they inform the user as to what regions of the parameter space new simulations (if needed) should be added
as well as how emulated and simulated data may differ in a high-dimensional feature space.
 Future work involves using these results 
 to design more efficient strategies for guided simulations
  that can balance statistical performance with computational costs.

\section*{Acknowledgements}
ABL, ND and TP were supported in part by the National Science Foundation under DMS-1520786.
RI was supported in part by FAPESP (2017/03363-8 and 2019/11321-9) and
CNPq (306943/2017-4).

%\newpage

\bibliographystyle{unsrt}
%\bibliography{paper}

\clearpage

\begin{center}
\textbf{\Large \bf Appendix: Validation of Approximate Likelihood and Emulator Models for Computationally Intensive Simulations}
\end{center}

\appendix 
%\annlee{11/29, The new text in the intro wasn't using bold-faced notation for $\x$. I fixed the notation in the intro and the main body of the text (I think), but I haven't made any updates yet to e.g. the proofs in Supp Mat.}
\section{Identifying Differences Between Two Multivariate Distributions}

\label{sec:morphology}

\begin{figure*}[!ht]
\centering
\includegraphics[width=0.45\textwidth]{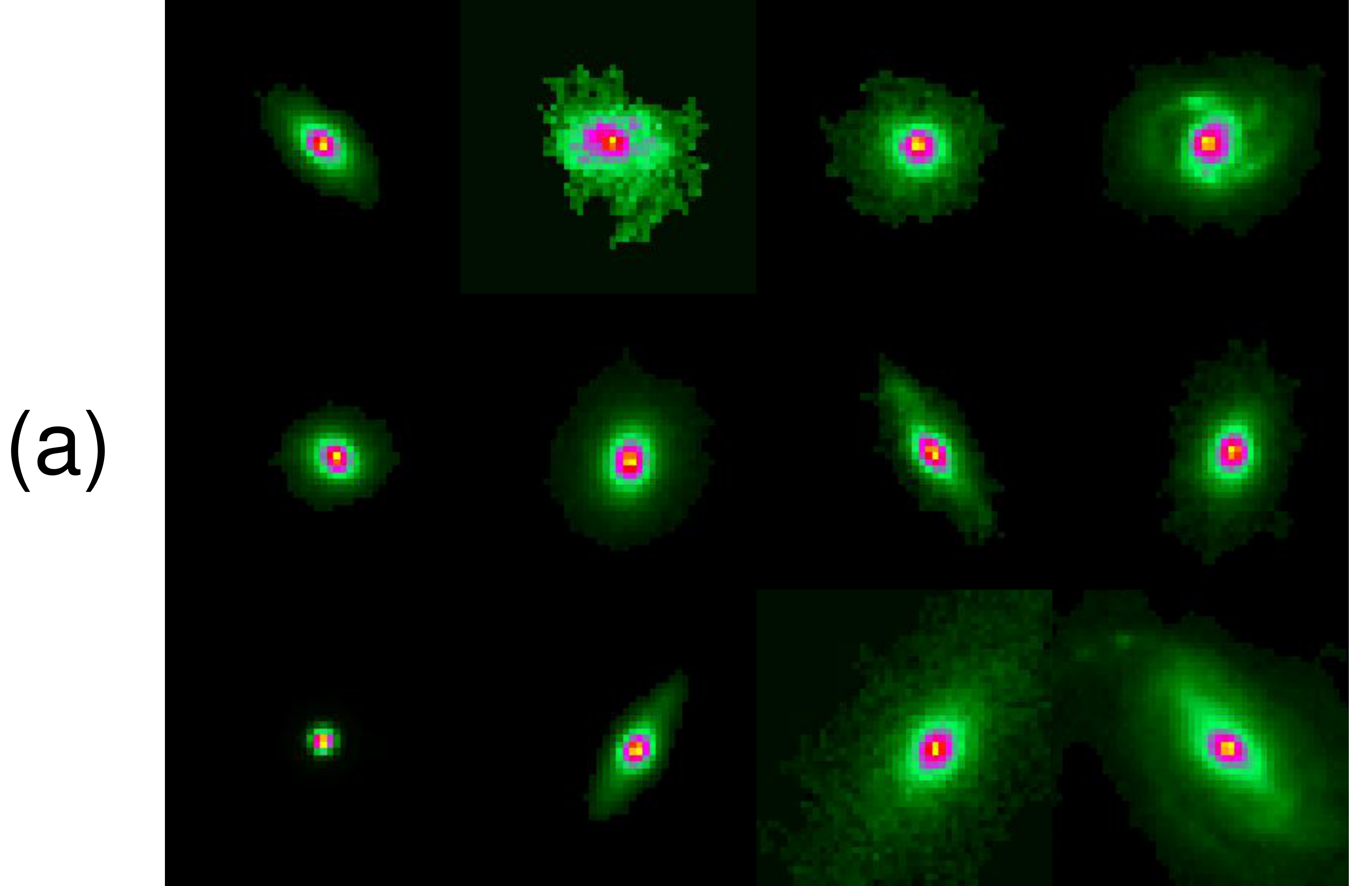} \hfill
\includegraphics[width=0.45\textwidth]{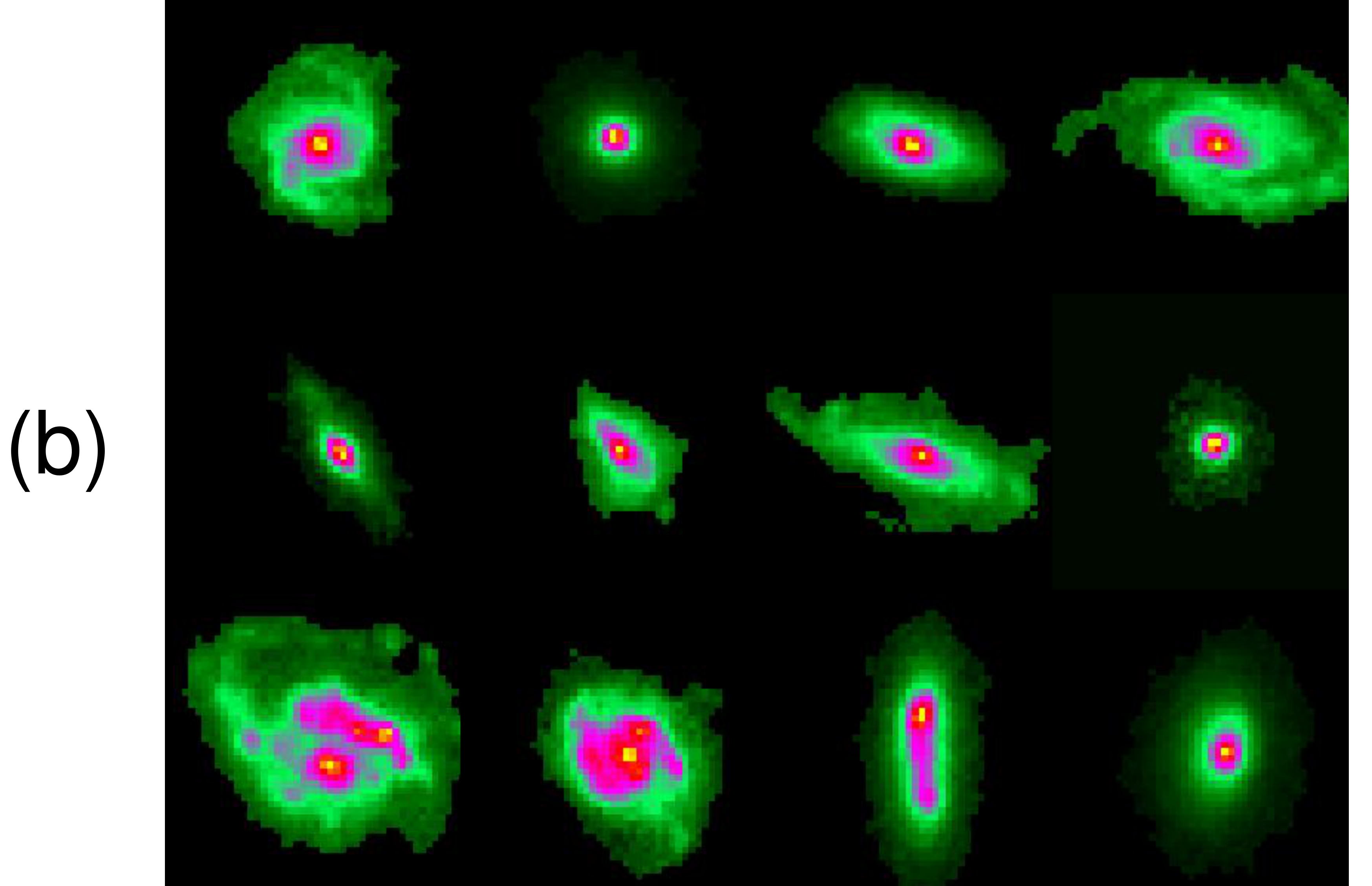}
\caption{\small{Examples of galaxies from (a) the low-SFR sample $\mathcal{S}_0$ versus (b) the high-SFR sample $\mathcal{S}_1$.}}
\label{fig: galaxy_samples}
\end{figure*}

\begin{figure*}[!ht]
\centering
\includegraphics[width=1\textwidth]{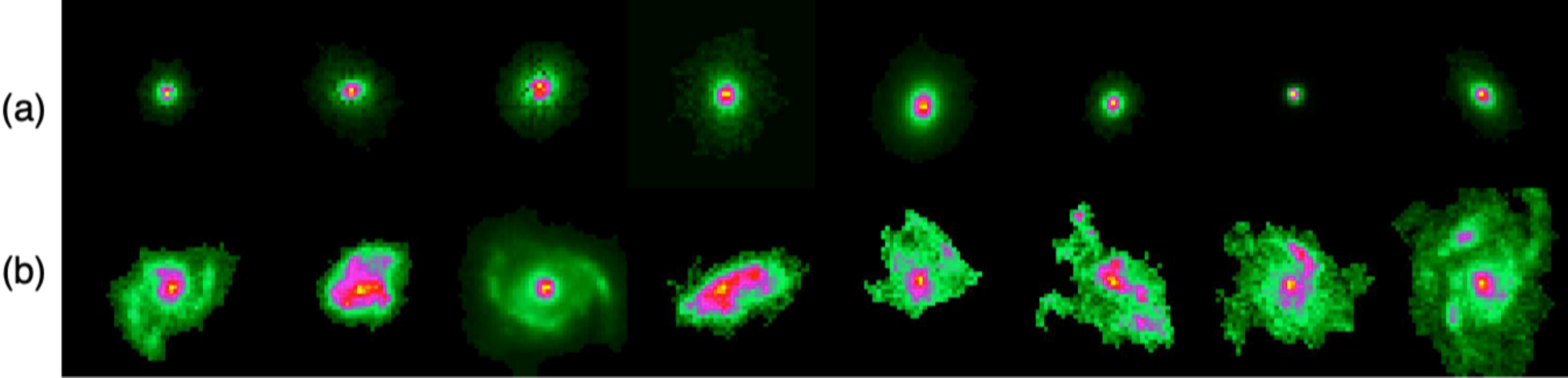}
\caption{\small{Galaxies in the test set with the highest significant difference $\left|\widehat{m}(\x) - \widehat{\pi}_1\right|$ according to our local test  in feature space, Algorithm \ref{alg:local_test_difference}. (a) Galaxies that are more representative of the low-SFR sample $\mathcal{S}_0$, and (b) galaxies more representative of the high-SFR sample  $\mathcal{S}_1$.  The first group of galaxies presents undisturbed and concentrated morphologies, while the latter galaxies appear more extended and/or disturbed. This is in line with what is expected by astronomers when comparing actual low-SFR and high-SFR galaxies.}}
\label{fig: sign_galaxies}
\end{figure*}

Here we show how our regression approach can be used to
identify and visualize locally significant differences between two multivariate distributions $P_0$ and $P_1$ defined over a ``feature space'' $\mathcal{X}$; we denote samples from the respective distributions by $\mathcal{S}_0$ and $\mathcal{S}_1$. In this example,  which is adapted from \cite{kim2016techreport}, $S_0$ and $S_1$ are both real observed data (galaxy images), but in our LFI setting $S_0$ would represent a sample from the simulator or target likelihood $\L(\x;\theta)$ at a fixed parameter value $\theta$, and $S_1$ would represent a sample from the emulator or approximate likelihood $\widehat{\L}(\x;\theta)$ (for the same parameter value). The goal would then be to identify with statistical confidence the regions in $\mathcal{X}$ which may be under- or over-represented in $\mathcal{S}_1$ (as compared to $S_0$). Our techniques can also be used to validate and diagnose output from generative adversarial networks (GANs) and other so-called implicit generative models
\cite{Mohamed2016LearningImplicitModels}; e.g., this type of analysis could be relevant for recent GAN models of galaxy images \citep{Ravanbakhsh2017GansGalaxies} and weak lensing convergence maps \citep{Mustafa2019cosmoGAN}.

\subsection*{Galaxy Morphology Example}

  \begin{algorithm}[!ht]
  \caption{ \small Local Test in Feature Space}\label{alg:local_test_difference}
  \algorithmicrequire \ {\small  
  i.i.d. training data from two populations $\{\X_i, Y_i\}_{i=1}^n$; testing data $\{\X_j\}_{j=1}^J$; number of permutations $M$; significance level $\alpha$; a regression method $\hat{m}$}\\ 
  \algorithmicensure \ {\small p-values $\{p_j\}_{j=1}^M$ for testing significance of difference $|\widehat{m}(\X_j) - \widehat{\pi}_1|$ for every test point}
  \begin{algorithmic}[1]
     \STATE $\widehat{\pi}_1 = 1/n \sum_{i=1}^n Y_i$;
     \STATE Train regression method $\hat{m}$ on training data $\{\X_i, Y_i\}_{i=1}^n$;
     \STATE Calculate the test statistics on each of the test points $$\hat{\nu}(\X_j) = (\hat{m}(\X_j) - \hat{\pi}_1)^2; $$
     \FOR{$k$ in $1,...,M$}
     \STATE Randomly permute $Y_1, ..., Y_n$ and train regression method on permuted data $\hat{m}^{(k)}$;
     \STATE Calculate the test statistics on the permuted data $\{\hat{\nu}^{(k)}(\X_j) = (\hat{m}^{(k)}(\X_j) - \hat{\pi}_1)^2\}_{j=1}^J; $
     \ENDFOR
     \STATE 
    \STATE Approximate permutation p-values $p_j$ for every test point $\X_j$:
    $$p_j = \frac{1}{M + 1} \sum_{k=1}^M \left( 1+  \mathbb{I} (\hat{\nu}^{(k)}(\X_j) > \hat{\nu}(\X_j)  ) \right)$$
    \STATE Apply a multiple test procedure to control false discovery rate;
    \STATE \textbf{return} $\{p_j\}_{j=1}^J$  
  \end{algorithmic}
  \end{algorithm}

Here we consider galaxies in the COSMOS, EGS, GOODS-North and UDS fields from CANDELS program \citep{grogin2011Candels, Koekemoer2011Candels}. The available data consist of seven morphology summary statistics 
 from $2736$ galaxies, together with their star formation rates (SFR).
We first sort the galaxies according to their star formation rates, and we define two populations --- with ``high'' SFR  ($Y=1$) versus ``low'' SFR ($Y=0$) --- by taking the top and bottom 25$^{\rm th}$ quantiles, respectively. Figure \ref{fig: galaxy_samples} shows a random subset of 12 galaxies from each sample. 

To compare the two populations in distribution, we use $65\%$ of the data to train a random forests regression, and the remaining $35\%$ for testing.
 For every test point $\x$ (that is, for every galaxy images in the test set), we compute the absolute difference 
$|\widehat{m}(\x) - \widehat{\pi}_1|$ between the estimated regression function and the proportion of high-SFR galaxies in the training sample.  We then calculate whether the difference $|\widehat{m}(\x) - \widehat{\pi}_1|$ is {\em statistically significant} according to a permutation test with a false discovery rate correction at $\alpha=0.05$ via Benjamini-Hochberg's method. The details of the local test in feature space are outlined in Algorithm \ref{alg:local_test_difference}.

\begin{figure*}[!ht]
\centering
\includegraphics[width=1\textwidth]{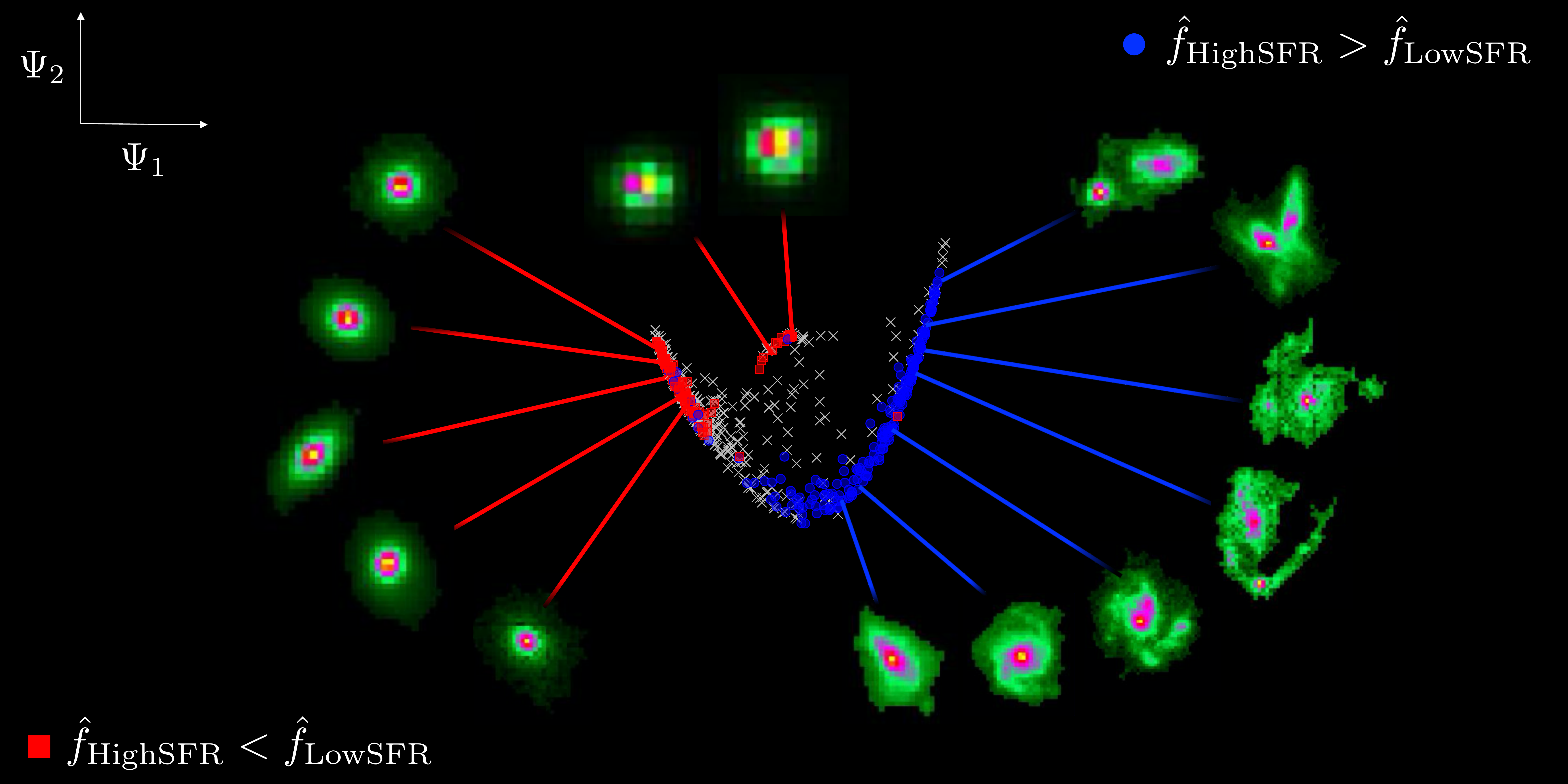}
\caption{\small  Results of two-sample testing of point-wise differences between high- and low-SFR galaxies in a seven-dimensional morphology space. The red color indicates regions where the density of low-SFR galaxies are significantly higher, and the blue color indicates regions that are dominated by high-SFR galaxies. The test points are visualized via a two-dimensional diffusion map. Figure adapted from \cite{kim2018regression}.}\label{fig: ilmun-sfr-galaxy}
\end{figure*}

Figure \ref{fig: sign_galaxies} shows examples of galaxies associated with the highest significant difference $|\widehat{m}(\x) - \widehat{\pi}_1|$; galaxies that are more representative of one sample than the other.
In Figure \ref{fig: ilmun-sfr-galaxy} we visualize the test data via a two-dimensional diffusion map \citep{Coifman2005DiffusionMaps}, where we color the test points that occur in regions of feature space where the local differences in the two distributions are  statistically significant.
 The blue points have $\widehat{m}(\x) > \widehat{\pi}_1$; these ``high-SFR regions'' are associated with extended, disturbed galaxy morphologies. The red points have $\widehat{m}(\x) < \widehat{\pi}_1$; these ``low-SFR regions'' are associated with concentrated, undisturbed morphologies. These results are consistent with what astronomers would expect, and illustrate the utility of the regression statistic $|\widehat{m}(\x) - \widehat{\pi}_1|$ in describing differences of two samples in a potentially high-dimensional feature space. For further details, see \citep{kim2016techreport, Freeman2017LocalTestGalaxy}.

\section{Proofs for the Global Test} \label{sec:proof_global_test}
\begin{Lemma}
\label{lemma:ks}
Let $\widehat{F}_{\mathbb{D}_{B,n_{\text{sim}}}}$ be the empirical cumulative distribution of the p-values in 
$\mathbb{D}_{B,n_{\text{sim}}}$,
$$\text{KS}(\mathbb{D}_{B,n_{\text{sim}}})=\sup_{0\leq z \leq 1}|\widehat{F}_{\mathbb{D}_{B,n_{\text{sim}}}}(z)-z|,$$
be the Kolmogorov-Smirnoff test statistic
and
$$\text{CVM}(\mathbb{D}_{B,n_{\text{sim}}})=\int_0^1 \left(\widehat{F}_{\mathbb{D}_{B,n_{\text{sim}}}}(z)-z\right)^2dz$$
be the Cram\'er-von Mises test statistic.
Both $\text{KS}$ and $\text{CVM}$ satisfy Assumptions \ref{assump:stat_zero}  and \ref{assump:uniformconsistent_stat}.
\end{Lemma}

\begin{proof}[\textbf{Proof of Lemma \ref{lemma:ks}}]
Let $U \sim U(0,1)$. From the law of large numbers,
\begin{align*}
    \text{KS}(\mathbb{U}_B)&=\sup_{0\leq z \leq 1}|\widehat{F}_{\mathbb{U}_B}(z)-z| \xrightarrow[B \longrightarrow \infty]{\text{a.s.}} \\
    &\quad \quad \quad \sup_{0\leq z \leq 1}|\P(U\leq z)-z|=0, 
\end{align*} 

which proves the first statement of the theorem.
Similarly, for every $n_{\text{sim}} \in \mathbb{N}$,
\begin{align}
\label{eq:s_conv}
    \text{KS}(\mathbb{D}_{B,n_{\text{sim}}}) &=
    \sup_{0\leq z \leq 1}|\widehat{F}_{\mathbb{D}_{B,n_{\text{sim}}}}(z)-z| 
    \xrightarrow[B \longrightarrow \infty]{\text{a.s.}} \nonumber \\
    &\quad \quad \quad
    \sup_{0\leq z \leq 1}|\P(p^{n_{\text{sim}}}_{\theta_1}\leq z)-z|.
\end{align}

Now,
Under Assumption \ref{assump:convergece_local},
for every $\theta_1 \in D$,
$$\P(p^{n_{\text{sim}}}_{\theta_1}\leq z|\theta_1) \xrightarrow{n_{\text{sim}} \longrightarrow \infty} 1$$
uniformly over $z \in (0,1)$.
Thus, under Assumption \ref{assump:support}, for every  $0<\epsilon_z <1-z$,
there exists
$n_{\text{sim}} \in \mathbb{N}$ such that, for every $n'_{\text{sim}}>n_{\text{sim}}$,
\begin{align}
\label{eq:cum}
\P(p^{n'_{\text{sim}}}_{\theta_1}\leq z)&=
\P(p^{n'_{\text{sim}}}_{\theta_1}\leq z|\theta_1 \in D)\P(\theta_1 \in D)
+ \notag \\
&\quad 
\P(p^{n'_{\text{sim}}}_{\theta_1}\leq z|\theta_1 \notin D)\P(\theta_1 \notin D)\notag \\ 
&\geq (1-\epsilon_z)\P(\theta_1 \in D)+
z \P(\theta_1 \notin D)\notag \\ 
&= (1-\epsilon_z+z-z)\P(\theta_1 \in D)+
z \P(\theta_1 \notin D)\notag \\ 
&= (1-\epsilon_z-z)\P(\theta_1 \in D)+ z\end{align}
It follows from Equations \ref{eq:s_conv} 
and \ref{eq:cum} and by taking $\epsilon_z=(1-z)/2$ that 
\begin{align*}
\sup_{0\leq z \leq 1}|\P(&p^{n'_{\text{sim}}}_{\theta_1}\leq z)-z|
\geq \sup_{0\leq z \leq 1}(1-\epsilon_z-z)\P(\theta_1 \in D) \\
&\geq \P(\theta_1 \in D)\sup_{0\leq z \leq 1}\frac{(1-z)}{2} =\frac{\P(\theta_1 \in D)}{2},
\end{align*}
and hence 
$$\lim_{n'_{\text{sim}} \longrightarrow \infty} \sup_{0\leq z \leq 1}|\P(p^{n'_{\text{sim}}}_{\theta_1}\leq z)-z| \geq \frac{\P(\theta_1 \in D)}{2}>0,$$
which concludes the proof for the KS statistic. The proof for the CVM statistic is analogous.
\end{proof}

\begin{proof}[\textbf{Proof of Theorem \ref{thm:main_control}}]
 Assumption \ref{assump:convergece_local}
 implies that $\phi_S$ is such that
$$\phi_S(\mathbb{D}_{B,n_{\text{sim}}})=1 \iff S(\mathbb{D}_{B,n_{\text{sim}}})\geq F^{-1}_{S(\mathbb{U}_{B})}(1-\alpha).$$
It follows that
\begin{align*}
&\P\left(\phi_{S}(\mathbb{D}_{B,n_{\text{sim}}})=1\right)  \\
& =  \P\left(S(\mathbb{D}_{B,n_{\text{sim}}})-F^{-1}_{S(\mathbb{U}_{B})}(1-\alpha) \geq 0\right) \\ 
&\geq \P\left(|S(\mathbb{D}_{B,n_{\text{sim}}})-a-F^{-1}_{S(\mathbb{U}_{B})}(1-\alpha)| \leq a\right) \\
& \quad \quad \xrightarrow{B,n_{\text{sim}} \longrightarrow \infty}1,
\end{align*}
where the last line follows from Assumptions \ref{assump:stat_zero}
and \ref{assump:uniformconsistent_stat}.
\end{proof}

\begin{proof}[\textbf{Proof of Corollary \ref{cor:ksVon}}]
 It follows directly from Theorem \ref{thm:main_control}
 and Lemma \ref{lemma:ks}.
\end{proof}

\section{Proofs for  Two-Sample Testing via Regression} \label{sec:proof_two_sample_regr}

\begin{Lemma} \label{Proposition: General Results of Global Regression Tests}
	Suppose that we have a regression estimate satisfying 
	\begin{align} \label{Eq: MSE_global}
\sup_{m \in \mathcal{M}} \mE \int_S \left( \widehat{m}(x) - m(x) \right)^2 dP_X(x) \leq C_0 \delta_n.
\end{align}
We reject the null hypothesis when $\widehat{\mathcal{T}}^\prime  \geq t_\alpha$ where $t_\alpha = 2 \max\{C_0, 1/4\} \alpha^{-1} \delta_n$. Then for any $\alpha,\beta \in (0,1/2)$, there exists a universal constant $C_1$ such that
	\begin{align*}
	& \text{$\bullet$ Type I error: } \mP_0 \left( \widehat{\mathcal{T}}^\prime  \geq t_\alpha \right) \leq \alpha \quad  \text{and} \\  
	& \text{$\bullet$ Type II error: } \sup_{m \in \mathcal{M}(C_1 \delta_n)} \mP_1 \left( \widehat{\mathcal{T}}^\prime  < t_\alpha \right) \leq \beta
	\end{align*}
	for a sufficiently large $n$. 
\end{Lemma}

	\begin{proof}[\textbf{Proof of Lemma \ref{Proposition: General Results of Global Regression Tests}}]
	We start with analyzing the type I error of the test.
	
	\vskip .8em 
	
	\noindent \textbf{$\bullet$ Type I Error Control}
	
	\noindent Under the null hypothesis, Markov's inequality shows that
	\begin{align*}
	&\mP_0 \left( \widehat{\mathcal{T}}^\prime  \geq t_\alpha \right) \leq \frac{\mE_0[\widehat{\mathcal{T}}^\prime ]}{t_\alpha} \\[.5em]
	&\quad \leq \frac{2}{t_\alpha} ( \mE_0 \left[  \int_S \left( \widehat{m}(x) - \pi_1 \right)^2 dP_X(x) \right] 
	+\mE_0 \left[ \left( \widehat{\pi}_1 - \pi_1 \right)^2  \right] ) \\[.5em]
	&\quad \leq \frac{2}{t_\alpha} \left( C_0 \delta_n  + \pi_1 (1-\pi_1) n^{-1} \right) \\
	&\quad \leq  \frac{2\max\{C_0, 1/4\}\delta_n}{t_\alpha}   = \alpha.
	\end{align*}
	Hence the result follows. Next, we control the type II error.
	
	\vskip .8em 
	
	\noindent \textbf{$\bullet$ Type II Error Control}

	\noindent Based on the inequality $(x-y)^2 \leq 2 (x-z)^2 + 2 (z-y)^2$, we lower bound the test statistic as 
	\begin{align} \nonumber
	\widehat{\mathcal{T}}_{}^\prime  & = \frac{1}{n} \sum_{i=n+1}^{2n} \left( \widehat{m}(X_i) - \widehat{\pi}_1 \right)^2 \\[.5em] \nonumber
	& \geq  \frac{1}{2n} \sum_{i=n+1}^{2n} \left( m(X_i) - \widehat{\pi}_1 \right)^2 \\
	&\quad - \frac{1}{n} \sum_{i=n+1}^{2n} \left( \widehat{m}(X_i) - m(X_i) \right)^2  \\[.5em] 
	& \geq \frac{1}{4n} \sum_{i=n+1}^{2n} \left( m(X_i) - \pi_1 \right)^2 - \frac{1}{2}(\pi_1 - \widehat{\pi}_1)^2 \nonumber \\
	&\quad - \frac{1}{n} \sum_{i=n+1}^{2n} \left( \widehat{m}(X_i) - m(X_i) \right)^2. \label{Eq: Lower Bound of Statistic}
	\end{align}
	Define the events $\mathcal{A}_1, \mathcal{A}_2, \mathcal{A}_3$ such that 
	\begin{align*}
	& \mathcal{A}_1 = \Big\{ (\pi_1 - \widehat{\pi}_1)^2 < C_2 \delta_n  \Big\}, \\[.5em]
	& \mathcal{A}_2 =  \Big\{  \frac{1}{n} \sum_{i=n+1}^{2n} \left( \widehat{m}(X_i) - m(X_i) \right)^2 < C_3\delta_n \Big\}, \\[.5em]
	&  \mathcal{A}_3 =  \Big\{  \Big|  \frac{1}{n} \sum_{i=n+1}^{2n} \left( m(X_i) - \pi_1 \right)^2 - \mE \left[ (m(X)- \pi_1)^2 \right] \Big| \\
	&\quad < \frac{1}{2} \mE \left[ (m(X)- \pi_1)^2 \right]  \Big\}.
	\end{align*}
	Using Markov's inequality, we have 
	\begin{align*}
	& \mP\left( \mathcal{A}_1^c  \right) \leq \frac{\pi_1(1-\pi_1)}{C_2 n \delta_n}, \\[.5em]
	& \mP\left( \mathcal{A}_2^c  \right) \leq \frac{1}{C_3 \delta_n} \mE\left[ \int_S (\widehat{m}(x) -m(x) )^2 dP_X(x) \right] \leq \frac{C_0}{C_3},
	\end{align*}
	by the condition in (\ref{Eq: MSE_global}). For the third event, denote $\Delta_n =  \mE\left[ (m(X) - \pi_1 )^2 \right]$ and use Chebyshev's inequality to have
	\begin{align*}
	\mP \left( \mathcal{A}_3^c \right) & \leq \frac{4}{n\Delta_n^2}  \text{Var} \left( (m(X) - \pi_1)^2 \right) \\[.5em]
	& \leq \frac{4}{n\Delta_n^2}  \mE \left[ (m(X) - \pi_1)^4 \right] \\[.5em]
	& \leq  \frac{4}{n\Delta_n^2}  \mE \left[ (m(X) - \pi_1)^2 \right]  \quad \text{since $|m(X) - \pi_1| \leq 1$ }\\[.5em]
	& \leq \frac{4}{C_1 n \delta_n},
	\end{align*}
	where the last inequality uses the assumption that $\Delta_n \geq C_1 \delta_n$. Hence, we obtain
	\begin{align*}
	\mP \left( (\mathcal{A}_1 \cap \mathcal{A}_2 \cap \mathcal{A}_3)^c \right)  \leq \mP \left(\mathcal{A}_1^c \right) + \mP \left(\mathcal{A}_2^c \right) + \mP \left(\mathcal{A}_3^c \right)  < \beta,
	\end{align*}
	by choosing sufficiently large $C_1,C_2,C_3 > 0$ with the assumption that $\delta_n \geq n^{-1}$. Using (\ref{Eq: Lower Bound of Statistic}), the type II error of the regression test is bounded by 
	\begin{align*}
	& \mP_1( \widehat{\mathcal{T}}_{}^\prime  < t_\alpha)  \\[.5em]
	\leq ~ & \mP_1 \Big(  \frac{1}{4n} \sum_{i=n+1}^{2n} \left( m(X_i) - \pi_1 \right)^2 - \frac{1}{2}(\pi_1 - \widehat{\pi}_1)^2 \\
	&\quad - \frac{1}{n} \sum_{i=n+1}^{2n} \left( \widehat{m}(X_i) - m(X_i) \right)^2 < t_\alpha \Big) \\[.5em]
	\leq ~ &  \mP_1 \Big(  \frac{1}{4n} \sum_{i=n+1}^{2n} \left( m(X_i) - \pi_1 \right)^2 - \frac{1}{2}(\pi_1 - \widehat{\pi}_1)^2 \\ 
	&\quad - \frac{1}{n} \sum_{i=n+1}^{2n} \left( \widehat{m}(X_i) - m(X_i) \right)^2 < t_\alpha,  \mathcal{A}_1 \cap \mathcal{A}_2 \cap \mathcal{A}_3 \Big) \\
	&\quad + \mP_1 \left((\mathcal{A}_1 \cap \mathcal{A}_2 \cap \mathcal{A}_3)^c \right) \\[.5em]
	\leq ~ & \mP_1 \left( \Delta_n < C_4 \delta_n \right) + \beta,
	\end{align*}
	where $C_4$ can be chosen by $C_4 = 4C_2 + 8C_3 + 16 \max \{ C_0, 1/4 \} /\alpha$. Now by choosing $C_1 > C_4$ for sufficiently large $n$, the type II error can be bounded by an arbitrary $\beta>0$. Hence the result follows.
\end{proof}

	\begin{proof}[\textbf{Proof of Theorem \ref{Theorem: Global Regression Test}}]
	The exact type I error control of the permutation test is well-known \citep[see e.g. Chapter 15 of][]{lehmann2006testing}. Hence we focus on the type II error control. 
	
	Let $\eta= (\eta_1,\ldots,\eta_{n})^\top$ be a permutation of $\{1,\ldots,{n}\}$. Now conditioned on the data $\mathcal{X}_{2n} = \{(X_1,Y_1),\ldots, (X_{2n},Y_{2n}) \}$, we denote the probability and expectation over permutations by $\mP_\eta[\cdot] = \mP_\eta[\cdot|\mathcal{X}_{2n}]$ and $\mE_\eta[\cdot] = \mE_\eta[\cdot|\mathcal{X}_{2n}]$ respectively. Then by Markov's inequality
	\begin{align*}
	\mP_\eta\left( \widehat{\mathcal{T}}^\prime  \geq t_\alpha^\ast \right) &= \mP_\eta \left( \frac{1}{n} \sum_{i=n+1}^{2n} \left( \widehat{m}_\eta(X_i) - \widehat{\pi}_1 \right)^2 \geq t_\alpha^\ast \right) \\
	&\quad \leq \frac{1}{t_\alpha^\ast n } \sum_{i=n+1}^{2n} \mE_\eta \left[ (\widehat{m}_\eta(X_i) - \widehat{\pi}_1)^2 \right], 
	\end{align*}
	where $\widehat{m}_\eta(x) = \sum_{i=1}^{n} w_i(x) Y_{\eta_i}$. Since $\sum_{i=1}^n w_i(x) = 1$ for any $x \in S$, 
	\begin{align*}
	\mE_\eta \left[ \widehat{m}_\eta (x) \right] = \sum_{i=1}^n w_i(x) \mE_{\eta} [Y_{\eta_i}] = \sum_{i=1}^n w_i(x) \widehat{\pi}_1 = \widehat{\pi}_1.
	\end{align*}
	Further note that 
		\begin{align}  \nonumber
	&\mE_\eta \left[ (\widehat{m}_\eta(x) - \widehat{\pi}_1)^2 \right] = \\ &\quad \quad \sum_{i_1=1}^{n} \sum_{i_2=1}^{n} w_{i_1}(x) w_{i_2}(x) \mE_{\eta} \left[ (Y_{\eta_{i_1}} - \widehat{\pi}_1 ) (Y_{\eta_{i_2}} - \widehat{\pi}_1 )\right] \\[.5em] \nonumber
	&\quad \leq \sum_{i=1}^n w_i^2(x) \mE_{\eta} \left[ (Y_{\eta_i} - \widehat{\pi}_1)^2 \right] \\[.5em] \nonumber
	&\quad = \widehat{\pi}_1 (1 - \widehat{\pi}_1 ) \sum_{i=1}^n w_i^2(x)  \leq \frac{1}{4} \sum_{i=1}^n w_i^2(x), \label{Eq: Upper bound of permutation moment}
	\end{align} 
	where the first inequality uses 
	$ \mE_{\eta} \left[ (Y_{\eta_{i_1}} - \widehat{\pi}_1 ) (Y_{\eta_{i_2}} - \widehat{\pi}_1 )\right]  \leq 0$ when $i_1 \neq i_2$.

	Note that the permutation samples are not $i.i.d.$ and thus in order to use the condition in (\ref{Eq: MSE_global}) which holds for $i.i.d.$ samples, we will associate the upper bound in (\ref{Eq: Upper bound of permutation moment}) with $i.i.d.$ samples. To do so,	let $(Y_1^\ast,\ldots, Y_n^\ast)$ be $i.i.d.$ Bernoulli random variables with parameter $p = 1/2$ independent of $\{X_1,\ldots,X_{2n}\}$. Then 
	\begin{align*}
	& \mE_{Y^\ast} \left[ (\widehat{m}(x) - 1/2)^2 | X_1, \ldots, X_{2n} \right] \\[.5em]
	= ~ &   \mE_{Y^\ast} \Big[ \big(\sum_{i=1}^n w_i(x) Y_i^\ast - 1/2 \big)^2 \big| X_1, \ldots, X_{2n} \Big]  \\[.5em]
	= ~ &   \mE_{Y^\ast} \Big[ \big(\sum_{i=1}^n w_i(x) (Y_i^\ast - 1/2 ) \big)^2 \big| X_1, \ldots, X_{2n} \Big]    \\[.5em]
	= ~ & \sum_{i_1=1}^{n} \sum_{i_2=1}^{n} w_{i_1}(x) w_{i_2}(x) \mE_{Y^\ast} [ (Y_{i_1}^\ast - 1/2) (Y_{i_2}^\ast - 1/2) ] \\[.5em]
	= ~ & \frac{1}{4} \sum_{i=1}^n w_i^2(x). 
	\end{align*}
	Therefore, we obtain
	\begin{align*}
	\mE_\eta \left[ (\widehat{m}_\eta(x) - \widehat{\pi}_1)^2 \right] & \leq ~ \mE_{Y^\ast} \left[ (\widehat{m}(x) - 1/2)^2 | X_1,\ldots,X_{2n} \right]
	\end{align*}
	which in turn implies that
	\begin{align*}
	&\mP_\eta \left( \widehat{\mathcal{T}}^\prime  \geq 
	t_\alpha^\ast \right)  \leq \\
	& \leq \frac{1}{t_\alpha^\ast n } \sum_{i=n+1}^{2n}\mE_{Y^\ast} \left[ (\widehat{m}(X_i) - 1/2 )^2 | X_1,\ldots,X_{2n} \right]. 
	\end{align*}
	So the critical value of the permutation distribution is bounded by 
	\begin{align*}
	t_{\alpha}^\ast \leq  \frac{1}{ \alpha n} \sum_{i=n+1}^{2n}\mE_{Y^\ast} \left[ (\widehat{m}(X_i) - 1/2 )^2 | X_1,\ldots,X_{2n} \right].
	\end{align*}
	
	Next, define the event 
	\begin{align}
	\mathcal{A} = \Bigg\{  \frac{1}{n} \sum_{i=n+1}^{2n}&\mE_{Y^\ast} \left[ (\widehat{m}(X_i) - 1/2 )^2 | X_1,\ldots,X_{2n} \right] \nonumber \\
	&\leq C_2^\prime \delta_n  \Bigg\}.
	\end{align}
	Now, because we assume that
	\begin{align} %\label{Eq: MSE_assumption}
\sup_{m \in \mathcal{M}} \mE \int_S \left( \widehat{m}(x) - m(x) \right)^2 dP_X(x) \leq C_0 \delta_n,
\end{align}
by Markov's inequality
it holds that
	\begin{align*}
	&\mP \left( \mathcal{A}^c \right)\\ 
	& \quad \leq 
	 \mP \Big(  \frac{1}{n} \sum_{i=n+1}^{2n}\mE_{Y^\ast} \left[ (\widehat{m}(X_i) - 1/2 )^2 | X_1,\ldots,X_{2n} \right] \\
	& \hspace{2cm} > C_2^\prime \delta_n \Big) \leq \frac{C_0}{C_2^\prime}.
	\end{align*}
	
	As a result, the type II error of the permutation test is bounded by
	\begin{align*}
	\mP_1 \left(\widehat{\mathcal{T}}^\prime  < t_\alpha^\ast \right) & \leq ~ \mP_1 \left(\widehat{\mathcal{T}}^\prime  < t_\alpha^\ast, \mathcal{A} \right) + \mP_1 \left( \mathcal{A}^c \right)  \\ \leq
	&\quad \mP_1 \left( \widehat{\mathcal{T}}^\prime  < \frac{C_2^\prime}{\alpha}\delta_n \right) + \frac{C_0}{C_2^\prime}.
	\end{align*}
	Now we choose $C_2^\prime$ sufficiently large so that
	\begin{align*}
	\frac{C_0}{C_2^\prime} < \frac{\beta}{2}.
	\end{align*}
	Next we follow the proof of Lemma~\ref{Proposition: General Results of Global Regression Tests} to show that 
	\begin{align*}
	\mP_1 \left( \widehat{\mathcal{T}}^\prime  < \frac{C_2^\prime}{\alpha}\delta_n  \right) < \frac{\beta}{2},
	\end{align*}
	which completes the proof. 
\end{proof}

\section{Goodness-of-Fit Regression Test via Monte Carlo Sampling}
\label{sec:test_MC_sampling}

 If the total number of test simulations from $\L(\x;\theta_0)$ is small, but the cost of drawing samples from the emulator model $\widehat{\L}(\x;\theta_0)$ is negligible, then we can instead of a two-sample permutation test perform a goodness-of-fit test, where we draw several independent Monte Carlo (MC) samples %$\{S^{(m)}\}_{m=1}^{M}$
 of size $n_e$ 
from $\widehat{\L}(\x;\theta_0)$ to produce a set of values $\{\widehat{T}^{(m)}\}_{m=1}^{M}$ that are used as a null distribution to test the hypothesis  $\L(\x;\theta_0)=\widehat{\L}(\x;\theta_0)$. (See Algorithm~\ref{alg:monte_carlo_regression_test} for details; here $f(\x)$ denotes the likelihood $\L(\x;\theta_0)$ of the simulator at $\theta=\theta_0$, and $f_e(\x)$ denotes the approximate likelihood  $\widehat{\L}(\x;\theta_0)$ of the emulator at the same parameter value.) If the emulations are cheap, we can choose $n_e \gg n_{\text{sim}}$ as well as a large number M. To cite Friedman \citep[Section IV]{friedman2004multivariate}, the goodness-of-fit approach has  ``the potential for increased power [compared to two-sample testing] at the expense of having to generate many Monte Carlo samples, instead of just one''.

Corollary \ref{cor:mc_sampling} states that our main result
 (Theorem \ref{Theorem: Global Regression Test}) still holds for the repeated MC sampling scheme. To simplify the proof, we again use sample splitting  for fitting the regression versus computing the test statistic.

 \begin{Cor} \label{cor:mc_sampling}
	Suppose that the regression estimator $\widehat{m}(\cdot)$ satisfies 
	\begin{align} \label{condition}
	\sup_{m \in \mathcal{M}} \mE \int_{\mathcal{X}} (\widehat{m}(\bx) - m(\bx))^2 dP_X(\bx) \leq C_0 \delta_n,
	\end{align}
	where $C_0$ is a positive constant, $\delta_n = o(1)$, $\delta_n \geq n^{-1}$ and $\mathcal{M}$ is a class of regression $m(\bx)$ containing constant functions. Given $M$ such that $\alpha > (M+1)^{-1}$, let us define the test via Monte Carlo sampling by
	\begin{align*}
	\phi_{\text{\emph{MC}}} = I\Bigg\{ \frac{1}{M+1} \left( 1 + \sum_{i=1}^M I ( \widehat{\mathcal{T}}_{\text{\emph{split}}}^{(i)} > \widehat{\mathcal{T}}_{\text{\emph{split}}}  ) \right) \leq \alpha \Bigg\}.
	\end{align*}
	Then for fixed $\alpha\in (0,1)$ and $\beta \in (1-\alpha)$ and sufficiently large $n_{\text{sim}}$ and $n_e$, there exists a constant $C_1$ such that 
	\begin{align*}
	\text{Type I error:} ~ & \mP_0(\phi_{\text{\emph{MC}}} = 1) \leq \alpha, \\[.5em]
	\text{Type II error:} ~ & \sup_{m \in \mathcal{M}(C_1\delta_n)}\mP_1(\phi_{\text{\emph{MC}}} = 0) \leq \beta,
	\end{align*} 
	against the class of alternatives $\mathcal{M}(C_1\delta_n) = \big\{ m \in \mathcal{M} : \int_{\mathcal{X}} (m(\bx) - \pi_1)^2 dP_X(\bx) \geq C_1 \delta_n \big\}$.
\end{Cor}

\textbf{Remark.} Here in contrast to the permutation approach, we do not assume that the regression is a linear smoother. 

\subsection{Proof of Corollary~\ref{cor:mc_sampling}}

We first prove the type I error control and then turn to the type II error control. 

\vskip 1em 

\noindent \textbf{$\bullet$ Type I error.}

\vskip .5em

\noindent With slight abuse of notation, let us write 
\begin{align*}
\phi_{\text{MC}}(\mathcal{T}) = I\Bigg\{ \frac{1}{M+1} \left( 1 + \sum_{i=1}^M I ( \widehat{\mathcal{T}}_{\text{{split}}}^{(i)} > \mathcal{T}) \right) \leq \alpha \Bigg\},
\end{align*}
so that $\phi_{\text{MC}}(\widehat{\mathcal{T}}_{\text{split}}) = \phi_{\text{MC}}$. By construction, it can be checked that 
\begin{align*}
\frac{1}{M} \sum_{i=1}^M \phi_{\text{MC}} (\widehat{\mathcal{T}}_{\text{{split}}}^{(i)}) \leq \alpha.
\end{align*}
Furthermore we know that $\widehat{\mathcal{T}}_{\text{split}}$ is equal in distribution to $\widehat{\mathcal{T}}_{\text{split}}^{(i)}$ for any $i=1,\ldots,M$ under the null hypothesis. Thus
\begin{align*}
\frac{1}{M} \sum_{i=1}^M \mE_{0} [\phi_{\text{MC}} (\widehat{\mathcal{T}}_{\text{{split}}}^{(i)})]= \mE_{0}[\phi_{\text{MC}}] \leq \alpha,
\end{align*} 
which verifies the type I error control.

\vskip 1em

\noindent \textbf{$\bullet$ Type II error.}

\vskip .5em

\noindent For this part of the proof, we closely follow the proof of Theorem 2.2 in \cite{kim2018regression}. We let denote the empirical distribution of Monte Carlo samples $\widehat{\mathcal{T}}^{(1)},\ldots,\widehat{\mathcal{T}}^{(M)}$ by
\begin{align*}
F_M(t) = \frac{1}{M} \sum_{i=1}^M I (\widehat{\mathcal{T}}^{(i)} \leq t) \quad \text{for all $t \in \mathbb{R}$.}
\end{align*} 
Then, by letting $ \alpha_M = \alpha (M+1)/M - 1/M$, we can see that $\phi_{\text{MC}} = 1$ if and only if $F_M(t) \geq 1 - \alpha_M$. In other words, we reject the null hypothesis if and only if 
\begin{align*}
\widehat{\mathcal{T}}_{\text{split}} \geq c_{1 - \alpha_M},
\end{align*}
where $c_{1 - \alpha_M}$ is the upper $1 - \alpha_M$ quantile of $F_M$. One can obtain an upper bound for this quantile by applying Markov's inequality as
\begin{align*}
c_{1 - \alpha_M} \leq \frac{1}{\alpha_M} \left( \frac{1}{M} \sum_{i=1}^M \widehat{\mathcal{T}}^{(i)}_{\text{split}} \right).
\end{align*}
Having this observation in mind and putting $\Delta_n = \mE[(m(\bX) - \pi_1)^2]$, let us define the events $\mathcal{A}_1, \mathcal{A}_2, \mathcal{A}_3$ such that
\begin{align*}
& \mathcal{A}_1 = \bigg\{ \frac{1}{M} \sum_{i=1}^M \widehat{\mathcal{T}}^{(i)}_{\text{split}}  \leq 3 \beta^{-1} C_0 \delta_n  \bigg\}, \\[.5em]
& \mathcal{A}_2 = \bigg\{ \frac{1}{n} \sum_{i=n+1}^{2n}( \widehat{m}(\mathbf{X}_i) - m(\mathbf{X}_i))^2 \leq 3\beta^{-1} C_0 \delta_n \bigg\} \quad \text{and} \\[.5em]
& \mathcal{A}_3 = \bigg\{ \bigg| \frac{1}{n} \sum_{i=n+1}^{2n}( m(\mathbf{X}_i) - \pi_1 )^2 - \Delta_n \bigg| \leq \Delta_n/2 \bigg\}.
\end{align*}
Then applying Markov's inequality together with condition~(\ref{condition}) yields $\mP(\mathcal{A}_1^c) \leq \beta/3$ and $\mP(\mathcal{A}_2^c) \leq \beta/3$. Moreover, as shown in \cite{kim2018regression}, we have $\mP(\mathcal{A}_3^c) \leq 4/(C_1n \delta_n)$. Combining these via the union bound, we see that the type II error is bounded by
\begin{align*}
\mP &\left(\widehat{\mathcal{T}}_{\text{split}} < c_{1 - \alpha_M} \right) \\[.5em]
& =  \mP \left(\widehat{\mathcal{T}}_{\text{split}} < c_{1 - \alpha_M}, \ \mathcal{A}_1 \right)  + \mP \left(\widehat{\mathcal{T}}_{\text{split}} < c_{1 - \alpha_M}, \ \mathcal{A}_1^c \right)  \\[.5em]
 & \leq  \mP \left(\widehat{\mathcal{T}}_{\text{split}} < 3 \alpha_M^{-1} \beta^{-1} C_0 \delta_n \right) + \mP(\mathcal{A}_1^c) \\[.5em]
& \leq   \mP \left(\widehat{\mathcal{T}}_{\text{split}} < 3 \alpha_M^{-1} \beta^{-1} C_0 \delta_n \right) + \frac{\beta}{3}.
\end{align*}
For the last line, based on the inequality $(x-y)^2 \leq 2 (x-z)^2 + 2(z-y)^2$, we further see that
\begin{align*}
& \mP \left(\widehat{\mathcal{T}}_{\text{split}} < 3 \alpha_M^{-1} \beta^{-1} C_0 \delta_n \right) \\[.5em] 
~ \leq ~ & \mP \bigg( \Big( \frac{1}{2n} \sum_{i=n+1}^{2n}( m(\mathbf{X}_i) - \pi_1 )^2 \\[0.5em]
- & \frac{1}{n} \sum_{i=n+1}^{n}( \widehat{m}(\mathbf{X}_i) - m(\bX_i) )^2 
 \Big) < 3 \alpha_M^{-1} \beta^{-1} C_0 \delta_n, \\[0.5em] 
 & \ \mathcal{A}_2 \cap \mathcal{A}_3 \Bigg)
+  \mP\left( \mathcal{A}_2^c \cup \mathcal{A}_3^c \right) \\[.5em]
\leq ~ & \mP\left( \Delta_n < 6(1+\alpha_M^{-1}) \beta^{-1} C_0 \delta_n \right) +  \frac{\beta}{3} + \frac{4}{C_1 n \delta_n}.
\end{align*}
Then by taking $C_1$ sufficiently large, the proof is complete.

\begin{algorithm}[!ht]
  \caption{ \small Goodness-of-Fit Regression Test via Monte Carlo Sampling}\label{alg:monte_carlo_regression_test}
  \algorithmicrequire \ {\small  i.i.d. sample $\mathcal{S}$ of size $n_{\text{sim}}$ from  distribution with density $f$;  emulator model with density $f_e$; size of Monte Carlo sample $n_e$; number of additional Monte Carlo samples $M$; a regression method $\widehat{m}$}\\ 
   \algorithmicensure \ {\small $p$-value for testing if $f(\x)=f_e(\x)$ for every $\x \in \mathcal{X}$} 
    \begin{algorithmic}[1]
     \STATE  Let $n=n_{\text{sim}}+n_e$.
    \STATE  Sample $\mathcal{S}_e=\{\X_1^{*},\ldots,\X_{n_e}^{*}\}$ from $f_e$.
    \STATE Define an augmented sample   $\{\X_i, Y_i\}_{i=1}^n$, where $\{\X_i\}_{i=1}^n \!\! = \!\! \mathcal{S}  \cup \mathcal{S}_e$, and $Y_i =I(\X_i \in  \mathcal{S}_e)$.
    \STATE Calculate the test statistic $\widehat{\mathcal T}$ in Equation~\ref{eq:regression_TS}.
     \FOR{$m \in \{1,\ldots,M\}$}
      \STATE Sample  $\mathcal{S}^{(m)}=\{\X_1^{(m)},\ldots,\X_{n_\text{sim}}^{(m)}\}$ from $f$, under the null hypothesis $H_0\!:\!f=f_e$.
     \STATE Sample  $\mathcal{S}^{(m)}_e=\{\X_1^{*(m)},\ldots,\X_{n_e}^{*(m)}\}$ from $f_e$. 
      \STATE Define a new augmented sample   $\{\X_i, Y_i\}_{i=1}^n$, where $\{\X_i\}_{i=1}^n \!\!= \!\! \mathcal{S}^{(m)}  \cup \mathcal{S}^{(m)}_e$, and $Y_i=I(\X_i \in  \mathcal{S}^{(m)}_e)$.
      \STATE Refit $\widehat{m}$ and calculate the test statistic on the new augmented sample to obtain  $\widehat{\mathcal T}^{(m)}$ from the null distribution $f=f_e$.
    \ENDFOR
     \STATE Compute the Monte Carlo $p$-value by
    		$p = \frac{1}{M+1} \left( 1 + \sum_{m=1}^M I(\widehat{\mathcal T}^{(m)} > \widehat{\mathcal T}  ) \right).$
     \STATE \textbf{return} $p$  
  \end{algorithmic}
\end{algorithm}

\section{Example 1 (Consistency of Global Test)}

In Example 1, we tested the null hypothesis that $\widehat{\mathcal{L}}(\x;\theta) =\mathcal{L}(\x;\theta)$ for data simulated according to $\theta \sim Gamma(1,1)$, and $\x=x_1,...,x_{1000}|\theta \sim Beta(\theta, \theta)$. Figure \ref{fig: likelihood_example1_parameter_grid} (left) shows the true likelihood $\mathcal{L}(\x;\theta)$ for some different values of $\theta$ \add{but a fixed $\x$ (for simplicity)}, comparing these functions to the likelihood approximation \add{$\widehat{\mathcal{L}}(\x;\theta) \propto 1$}. Such an approximation is valid when $\theta=1$, as $Beta(1,1)$ is indeed just the uniform distribution, whereas the approximation is clearly wrong for the other values of $\theta \sim Gamma(1,1)$.

\section{Example 2 (Power of Two-Sample Test via Regression)}
The practical implications of Theorem \ref{Theorem: Global Regression Test} are that for a two-sample test via regression one should base the test on the regression method with the  smallest mean integrated squared error (MISE) so as to achieve a more powerful test.
%to have the higher power. 
Table \ref{tab: ex2_supp_mat} illustrates this for the three settings in Example 2 (Section \ref{sec:examples_gof}): random forest achieves a smaller MISE than nearest neighbor (NN) regression across all settings and, as Figure \ref{fig::synthetic_examples} shows, it also consistently attains a higher power.
\begin{table}[!ht]
\begin{tabular}{|c||c|c|}
\hline
\textit{Setting / Regression Method} & \textit{Random Forest} & \textit{NN} \\ \hline
(a) Bernoulli                     & 0.19                   & 0.73                      \\ \hline
(b) Scaling                       & 0.35                   & 2.31                      \\ \hline
(c) Mixture of Gaussians          & 0.27                   & 1.64                      \\ \hline
\end{tabular}
\caption{\small{Integrated mean squared error (MISE) for regression methods used for two-sample testing in Figure \ref{fig::synthetic_examples}. Random forest has the smallest MISE in regression; it also yields the test with highest power, as implied by Theorem \ref{Theorem: Global Regression Test}.}}\label{tab: ex2_supp_mat}
\end{table}

\begin{figure}[!ht]
\centering
\includegraphics[width=0.5\textwidth]{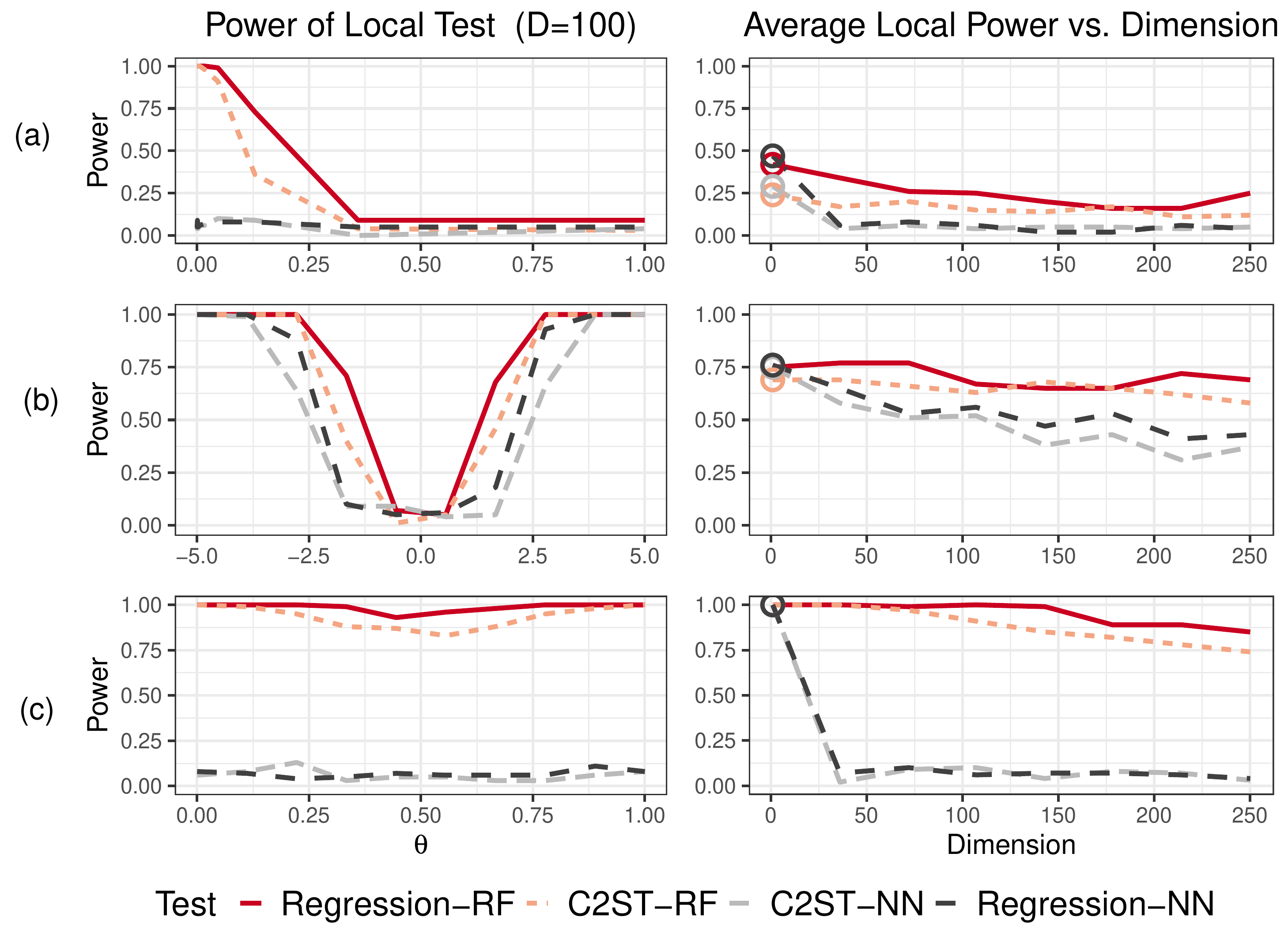}
\caption{\small  Test power at $D=100$ (left column) and as a function of dimension $D$ (right column) in the same Example 2 settings, i.e., for (a) Bernoulli, (b) Scaling and (c) Mixture of Gaussians. We include the results for our regression test with random forests (RF) and nearest neighbors (NN), as well as the corresponding results using the classification accuracy test of \cite{LopezPaz2017C2ST} with RF and NN (labeled as C2ST-RF and C2ST-NN, respectively).
} \label{fig::synthetic_examples_c2st_supp_mat} 
\end{figure}

As pointed out in the related work section, classifier two-sample testing methods have also been used for two-sample testing by dichotomizing the regression function and using the classification accuracy as a test statistic. Such dichotomization might result in a loss of power with respect to the respective regression test in certain settings (for more examples, see \cite{kim2018regression}). In Figure \ref{fig::synthetic_examples_c2st_supp_mat} we consider the same settings as in Example 2, but now also computing the power of the classification accuracy test from \cite{LopezPaz2017C2ST} for both random forest and nearest neighbor classification. The regression test achieves comparable results across the different settings, providing slight improvements in some cases, e.g., with respect to the local power at $D=100$ (left column). Note that our global procedure can incorporate classification accuracy tests as well, but would then not be able to identify locally significant differences in feature space as in Section \ref{sec:morphology}.

\section{Approximate P-Values and Confidence Regions} \label{sec:approx_pval_conf_regions}

Consider testing $H_0:\theta\in \Theta_0$. Let 
$\lambda(\x)$ be the likelihood ratio statistic for testing $H_0$, i.e.,
$$\lambda(\x)=\frac{\sup_{\theta \in \Theta_0}\mathcal{L}(\x;\theta)}{\sup_{\theta \in \Theta}\mathcal{L}(\x;\theta)}.$$
We estimate $\lambda(\x)$ using the estimated likelihood:
$$\widehat{\lambda}(\x)=\frac{\sup_{\theta \in \Theta_0}\widehat{\mathcal{L}}(\x;\theta)}{\sup_{\theta \in \Theta}\widehat{\mathcal{L}}(\x;\theta)}.$$
The estimated p-value is then
$$\widehat{p}(\x)=\sup_{\theta \in \Theta_0} \P_{\theta}(\widehat{\lambda}(\X)>\widehat{\lambda}(\x))$$ 
If $\Theta_0=\{\theta_0\}$,
$\widehat{p}(\x)$ can be estimated
using  data that are  simulated under $\theta=\theta_0$. 
If $|\Theta_0|>1$,
the distribution of 
the test statistic can be approximated using the $\chi^2$
approximation for the likelihood
ratio test \citep{casella2002statistical}.
Confidence intervals may be obtained by inverting the hypothesis tests \citep{casella2002statistical}.

\begin{figure*}[!ht]
\centering
\includegraphics[width=0.475\textwidth]{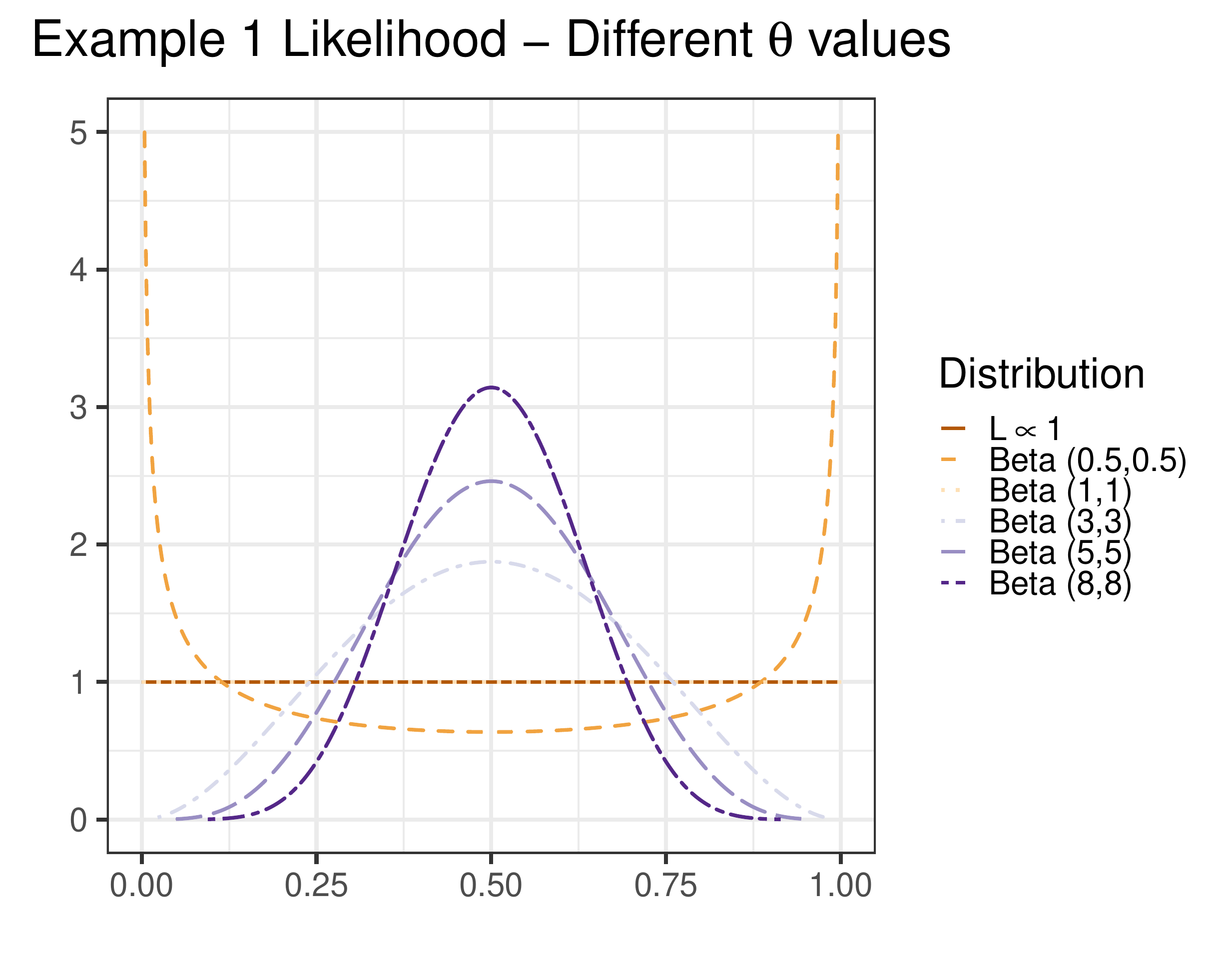}
\includegraphics[width=0.475\textwidth]{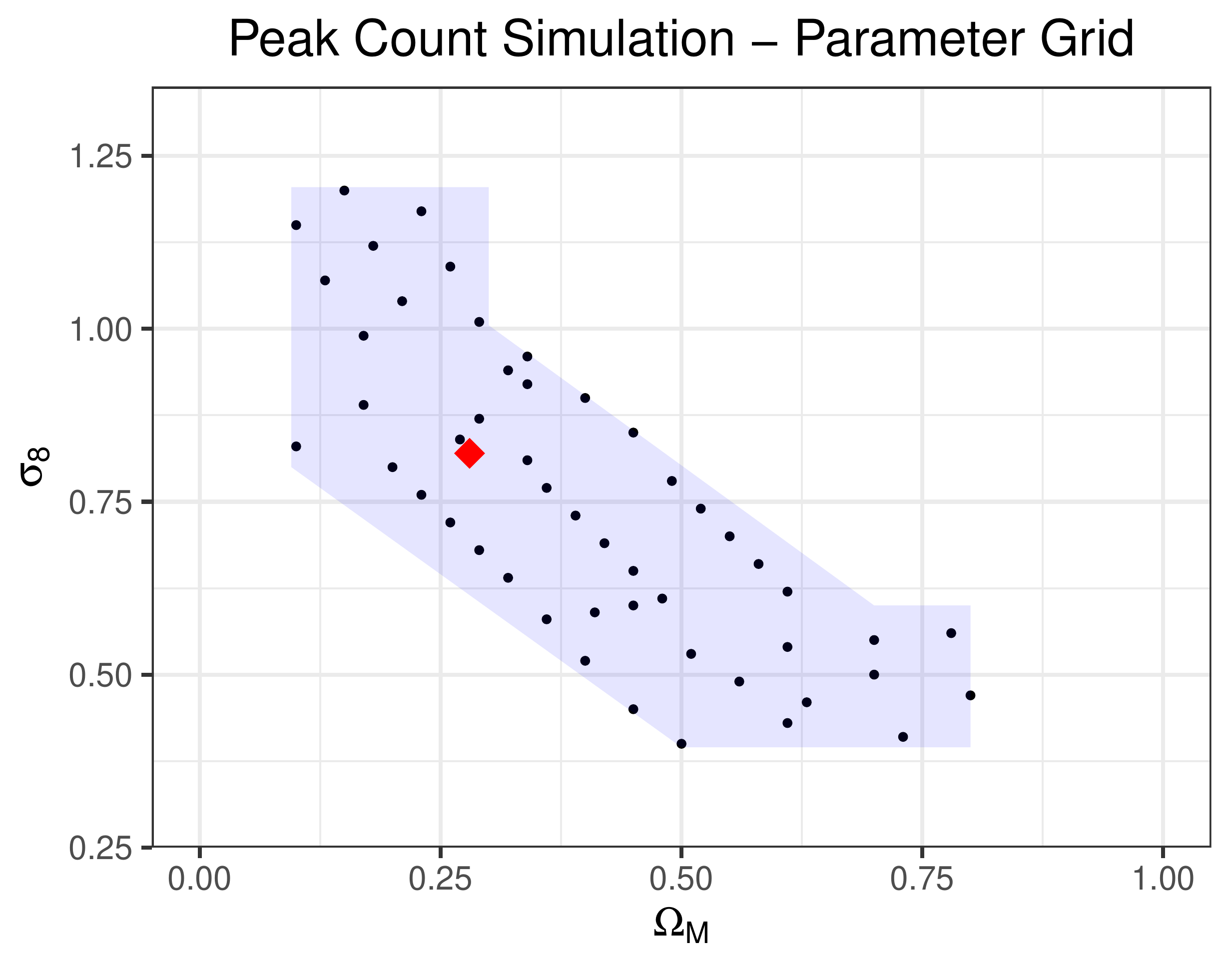}
\caption{\small {\em Left}: The true likelihood for different values of the parameter $\theta$, compared to the approximation $\widehat{\mathcal{L}}(\x; \theta) \propto 1$. The approximation is clearly wrong when $\theta \neq 1$. {\em Right}: Location of the 50 parameter settings for the peak count data simulations using \texttt{CAMELUS}, where the blue region indicates the parameter range values and the red diamond indicates the fiducial point $\theta_0$.}
\label{fig: likelihood_example1_parameter_grid}
\end{figure*}

\section{Peak Count Data Example}

The KL divergence for model comparison is estimated by:
\begin{align*}
KL(\mathcal{L},\widehat{\mathcal{L}})&=-\E\left[\log \left(\frac{\widehat{\mathcal{L}}(\x;\theta)}{\mathcal{L}(\x;\theta)}\right) \right]\\
&=-\E\left[\log \left(\widehat{\mathcal{L}}(\x;\theta)\right) \right]+K \\ &\approx -\frac{1}{n}  \sum_{j=1}^m \sum_{i=1}^{n_j} \log \left(\widehat{\mathcal{L}}(\x_{ij};\theta_j)\right) + K
\end{align*}
where $K$ does not depend on $\widehat{\mathcal{L}}$; $\{\theta_j\}_{j=1}^m$  with  $m=50$ denotes the parameters used by the simulator; $\{\x_{ij}\}_{i=1}^{n_j}$ (with $n_j=200$ for all $\theta_j$) denotes the test simulations at $\theta_j$; and $\sum_{j=1}^m n_j = n$ is the total number of test simulations.

Figure \ref{fig: likelihood_example1_parameter_grid}, right, shows the grid of 50 parameters settings $\theta = (\Omega_{m}, \sigma_8)$ which we use for the \texttt{CAMELUS} batch simulations. The blue shaded region represents the parameter regions from which the parameters are sampled around the fiducial cosmology $\theta_0$ (indicated by a red diamond).

For the conditional MAF, at both $n_{\text{train}}=200$ and $n_{\text{train}}=500$ we used $10\%$ of the training data as validation. During training we assessed validation loss and we stopped the training early if the validation loss was not improving for 30 epochs. We explored architectures with $\{5, 10, 15, 20\}$ autoregressive layers and $2^{\{4,..,10\}}$ hidden units, with the best performing having $10$ autoregressive layers and either $512$ or $1024$ hidden units.

%\end{document}

\end{document}